\documentclass[prd,twocolumn,showpacs,amssymb,amsmath,amsfonts,aps,superscriptaddress,altaffilletter]{revtex4}
\pdfoutput=1

\usepackage{color}
\usepackage{times}
\usepackage{graphicx}
\usepackage{fancyhdr}
\usepackage{float}
\usepackage{ulem}
\usepackage{acronym}
\usepackage{multirow}
\usepackage{array}
\usepackage{mathptmx}
\usepackage{booktabs}
\numberwithin{equation}{section}

\makeatletter
\makeatletter
\def\@fnsymbol#1{\ifcase#1\or * \or  $+$ \or  \$ \or \#  \or \dag \or \ddag \or
$\mathsection$ \or $ \mathparagraph$ \or $\|$  \or \textordfeminine \or \textbul
let   
\or ** \or $++$ \or  \$\$ \or \#\#  \or \dag\dag \or \ddag\ddag \or
$\mathsection\mathsection$ \or $ \mathparagraph\mathparagraph$ \or $\|\|$  \or 
\textordfeminine\textordfeminine \or \textbullet \textbullet \or *** \or $+++$ 
\or  \$\$\$ \or \#\#  \or \dag\dag \or \ddag\ddag \or
$\mathsection \mathsection\mathsection$ \or $ \mathparagraph 
\mathparagraph\mathparagraph$ \or $\|\|\|$  \or 
\textordfeminine\textordfeminine\textordfeminine \or 
\textbullet\textbullet\textbullet \or \else \@ctrerr\fi}
\makeatother

\newcommand\fake[1]{\textcolor{red}{#1}}

\newcommand{\diff}{{d}}
\def\thercsid{\relax}
\def\rcsid#1{\def\next##1#1{\def\thercsid{##1}}\next}

\rcsid$Id: s5-highMass.tex,v 1.164 2011/02/17 16:59:34 evano Exp $

\renewcommand{\today}{\number\day\space\ifcase\month\or
  January\or February\or March\or April\or May\or June\or
  July\or August\or September\or October\or November\or December\fi
  \space\number\year}

\def\Msun{\ensuremath{M_{\odot}}}

\newcommand{\un}[1]{\text{\,#1}}

\def\ulnum{\ensuremath{2.0 \un{Mpc}^{-3}\un{Myr}^{-1}}}
\def\ulrange{\ensuremath{19\,\Msun\le m_1,m_2\le 28\,\Msun}}
\def\FAR{\ensuremath{\mathrm{FAR}}}

\begin{document}

\acrodef{BH}{black hole}
\acrodef{BHs}{black holes}
\acrodef{BBHs}{binary black holes}
\acrodef{BBH}{binary black hole}
\acrodef{BNS}{binary neutron stars}
\acrodef{BHNS}{black hole-neutron star binaries}
\acrodef{PBH}{primordial black hole binaries}
\acrodef{SNR}{signal-to-noise ratio}
\acrodef{SPA}{stationary-phase approximation}
\acrodef{EOB}{effective one-body}
\acrodef{LIGO}{Laser Interferometer Gravitational-wave Observatory}
\acrodef{LHO}{LIGO Hanford Observatory}
\acrodef{LLO}{LIGO Livingston Observatory}
\acrodef{LSC}{LIGO Scientific Collaboration}
\acrodef{GRB}{gamma-ray bursts}
\acrodef{CBC}{compact binary coalescences}
\acrodef{GW}{gravitational wave}
\acrodef{ISCO}{innermost stable circular orbit}
\acrodef{FAR}{false alarm rate}
\acrodef{IFAR}{inverse false alarm rate}
\acrodef{CL}{confidence level}
\acrodef{PN}{post-Newtonian}
\acrodef{DQ}{data quality}
\acrodef{IMR}{inspiral, merger and ringdown}
\acrodef{NR}{numerical relativity}
\acrodef{S5}{fifth science run}

\title{Search for gravitational waves from binary black hole inspiral, merger and ringdown\\
}

\date[\relax]{Dated: \today }
\fake{}


\begin{abstract}\quad
\newpage
We present the first modeled search for gravitational waves using the complete
binary black hole gravitational waveform from inspiral through the merger and
ringdown for binaries with negligible component spin.  We searched
approximately 2 years of LIGO data taken between November 2005 and September
2007 for systems with component masses of 1--99\,{\Msun} and total masses of
25--100\,{\Msun}.  We did not detect any plausible gravitational-wave signals
but we do place upper limits on the merger rate of binary black holes as a
function of the component masses in this range.  We constrain the rate of
mergers for \ulrange binary black hole systems with negligible spin to be no
more than \ulnum at 90\% confidence.
\end{abstract}





\affiliation{Albert-Einstein-Institut, Max-Planck-Institut f\"ur Gravitationsphysik, D-14476 Golm, Germany$^\ast$}
\affiliation{Albert-Einstein-Institut, Max-Planck-Institut f\"ur Gravitationsphysik, D-30167 Hannover, Germany$^\ast$}
\affiliation{Andrews University, Berrien Springs, MI 49104 USA$^\ast$}
\affiliation{Laboratoire AstroParticule et Cosmologie (APC) Universit\'e Paris Diderot, CNRS: IN2P3, CEA: DSM/IRFU, Observatoire de Paris, 10 rue A.Domon et L.Duquet, 75013 Paris - France$^\dagger$}
\affiliation{Australian National University, Canberra, 0200, Australia$^\ast$}
\affiliation{California Institute of Technology, Pasadena, CA  91125, USA$^\ast$}
\affiliation{California State University Fullerton, Fullerton CA 92831 USA$^\ast$}
\affiliation{Caltech-CaRT, Pasadena, CA  91125, USA$^\ast$}
\affiliation{Cardiff University, Cardiff, CF24 3AA, United Kingdom$^\ast$}
\affiliation{Carleton College, Northfield, MN  55057, USA$^\ast$}
\affiliation{Charles Sturt University, Wagga Wagga, NSW 2678, Australia$^\ast$}
\affiliation{Columbia University, New York, NY  10027, USA$^\ast$}
\affiliation{European Gravitational Observatory (EGO), I-56021 Cascina (PI), Italy$^\dagger$}
\affiliation{Embry-Riddle Aeronautical University, Prescott, AZ   86301 USA$^\ast$}
\affiliation{E\"otv\"os Lor\'and University, Budapest, 1117 Hungary$^\ast$}
\affiliation{Hobart and William Smith Colleges, Geneva, NY  14456, USA$^\ast$}
\affiliation{INFN, Sezione di Firenze, I-50019 Sesto Fiorentino$^a$; Universit\`a degli Studi di Urbino 'Carlo Bo', I-61029 Urbino$^b$, Italy$^\dagger$}
\affiliation{INFN, Sezione di Genova;  I-16146  Genova, Italy$^\dagger$}
\affiliation{INFN, Sezione di Napoli $^a$; Universit\`a di Napoli 'Federico II'$^b$ Complesso Universitario di Monte S.Angelo, I-80126 Napoli; Universit\`a di Salerno, Fisciano, I-84084 Salerno$^c$, Italy$^\dagger$}
\affiliation{INFN, Sezione di Perugia$^a$; Universit\`a di Perugia$^b$, I-06123 Perugia,Italy$^\dagger$}
\affiliation{INFN, Sezione di Pisa$^a$; Universit\`a di Pisa$^b$; I-56127 Pisa; Universit\`a di Siena, I-53100 Siena$^c$, Italy$^\dagger$}
\affiliation{INFN, Sezione di Roma$^a$; Universit\`a 'La Sapienza'$^b$, I-00185 Roma, Italy$^\dagger$}
\affiliation{INFN, Sezione di Roma Tor Vergata$^a$; Universit\`a di Roma Tor Vergata, I-00133 Roma$^b$; Universit\`a dell'Aquila, I-67100 L'Aquila$^c$, Italy$^\dagger$}
\affiliation{Institute of Applied Physics, Nizhny Novgorod, 603950, Russia$^\ast$}
\affiliation{Inter-University Centre for Astronomy and Astrophysics, Pune - 411007, India$^\ast$}
\affiliation{LAL, Universit\'e Paris-Sud, IN2P3/CNRS, F-91898 Orsay$^a$; ESPCI, CNRS,  F-75005 Paris$^b$, France$^\dagger$}
\affiliation{Laboratoire d'Annecy-le-Vieux de Physique des Particules (LAPP), Universit\'e de Savoie, CNRS/IN2P3, F-74941 Annecy-Le-Vieux, France$^\dagger$}
\affiliation{Leibniz Universit\"at Hannover, D-30167 Hannover, Germany$^\ast$}
\affiliation{LIGO - California Institute of Technology, Pasadena, CA  91125, USA$^\ast$}
\affiliation{LIGO - Hanford Observatory, Richland, WA  99352, USA$^\ast$}
\affiliation{LIGO - Livingston Observatory, Livingston, LA  70754, USA$^\ast$}
\affiliation{LIGO - Massachusetts Institute of Technology, Cambridge, MA 02139, USA$^\ast$}
\affiliation{Laboratoire des Mat\'eriaux Avanc\'es (LMA), IN2P3/CNRS, F-69622 Villeurbanne, Lyon, France$^\dagger$}
\affiliation{Louisiana State University, Baton Rouge, LA  70803, USA$^\ast$}
\affiliation{Louisiana Tech University, Ruston, LA  71272, USA$^\ast$}
\affiliation{McNeese State University, Lake Charles, LA 70609 USA$^\ast$}
\affiliation{Montana State University, Bozeman, MT 59717, USA$^\ast$}
\affiliation{Moscow State University, Moscow, 119992, Russia$^\ast$}
\affiliation{NASA/Goddard Space Flight Center, Greenbelt, MD  20771, USA$^\ast$}
\affiliation{National Astronomical Observatory of Japan, Tokyo  181-8588, Japan$^\ast$}
\affiliation{Nikhef, Science Park, Amsterdam, the Netherlands$^a$; VU University Amsterdam, De Boelelaan 1081, 1081 HV Amsterdam, the Netherlands$^b$$^\dagger$}
\affiliation{Northwestern University, Evanston, IL  60208, USA$^\ast$}
\affiliation{Universit\'e Nice-Sophia-Antipolis, CNRS, Observatoire de la C\^ote d'Azur, F-06304 Nice$^a$; Institut de Physique de Rennes, CNRS, Universit\'e de Rennes 1, 35042 Rennes$^b$, France$^\dagger$}
\affiliation{INFN, Gruppo Collegato di Trento$^a$ and Universit\`a di Trento$^b$,  I-38050 Povo, Trento, Italy;   INFN, Sezione di Padova$^c$ and Universit\`a di Padova$^d$, I-35131 Padova, Italy$^\dagger$}
\affiliation{IM-PAN 00-956 Warsaw$^a$; Warsaw University 00-681 Warsaw$^b$; Astronomical Observatory Warsaw University 00-478 Warsaw$^c$; CAMK-PAN 00-716 Warsaw$^d$; Bia{\l$^\ast$}ystok University 15-424 Bia{\l$^\ast$}ystok$^e$; IPJ 05-400 \'Swierk-Otwock$^f$; Institute of Astronomy 65-265 Zielona G\'ora$^g$,  Poland$^\dagger$}
\affiliation{RMKI, H-1121 Budapest, Konkoly Thege Mikl\'os \'ut 29-33, Hungary$^\dagger$}
\affiliation{Rochester Institute of Technology, Rochester, NY  14623, USA$^\ast$}
\affiliation{Rutherford Appleton Laboratory, HSIC, Chilton, Didcot, Oxon OX11 0QX United Kingdom$^\ast$}
\affiliation{San Jose State University, San Jose, CA 95192, USA$^\ast$}
\affiliation{Sonoma State University, Rohnert Park, CA 94928, USA$^\ast$}
\affiliation{Southeastern Louisiana University, Hammond, LA  70402, USA$^\ast$}
\affiliation{Southern University and A\&M College, Baton Rouge, LA  70813, USA$^\ast$}
\affiliation{Stanford University, Stanford, CA  94305, USA$^\ast$}
\affiliation{Syracuse University, Syracuse, NY  13244, USA$^\ast$}
\affiliation{The Pennsylvania State University, University Park, PA  16802, USA$^\ast$}
\affiliation{The University of Melbourne, Parkville VIC 3010, Australia$^\ast$}
\affiliation{The University of Mississippi, University, MS 38677, USA$^\ast$}
\affiliation{The University of Sheffield, Sheffield S10 2TN, United Kingdom$^\ast$}
\affiliation{The University of Texas at Austin, Austin, TX 78712, USA$^\ast$}
\affiliation{The University of Texas at Brownsville and Texas Southmost College, Brownsville, TX  78520, USA$^\ast$}
\affiliation{Trinity University, San Antonio, TX  78212, USA$^\ast$}
\affiliation{Tsinghua University, Beijing 100084 China$^\ast$}
\affiliation{Universitat de les Illes Balears, E-07122 Palma de Mallorca, Spain$^\ast$}
\affiliation{University of Adelaide, Adelaide, SA 5005, Australia$^\ast$}
\affiliation{University of Birmingham, Birmingham, B15 2TT, United Kingdom$^\ast$}
\affiliation{University of Florida, Gainesville, FL  32611, USA$^\ast$}
\affiliation{University of Glasgow, Glasgow, G12 8QQ, United Kingdom$^\ast$}
\affiliation{University of Maryland, College Park, MD 20742 USA$^\ast$}
\affiliation{University of Massachusetts - Amherst, Amherst, MA 01003, USA$^\ast$}
\affiliation{University of Michigan, Ann Arbor, MI  48109, USA$^\ast$}
\affiliation{University of Minnesota, Minneapolis, MN 55455, USA$^\ast$}
\affiliation{University of Oregon, Eugene, OR  97403, USA$^\ast$}
\affiliation{University of Rochester, Rochester, NY  14627, USA$^\ast$}
\affiliation{University of Salerno, I-84084 Fisciano (Salerno), Italy and INFN (Sezione di Napoli), Italy$^\ast$}
\affiliation{University of Sannio at Benevento, I-82100 Benevento, Italy and INFN (Sezione di Napoli), Italy$^\ast$}
\affiliation{University of Southampton, Southampton, SO17 1BJ, United Kingdom$^\ast$}
\affiliation{University of Strathclyde, Glasgow, G1 1XQ, United Kingdom$^\ast$}
\affiliation{University of Western Australia, Crawley, WA 6009, Australia$^\ast$}
\affiliation{University of Wisconsin--Milwaukee, Milwaukee, WI  53201, USA$^\ast$}
\affiliation{Washington State University, Pullman, WA 99164, USA$^\ast$}
\author{J.~Abadie$^\text{29}$}\noaffiliation\author{B.~P.~Abbott$^\text{29}$}\noaffiliation\author{R.~Abbott$^\text{29}$}\noaffiliation\author{M.~Abernathy$^\text{67}$}\noaffiliation\author{T.~Accadia$^\text{27}$}\noaffiliation\author{F.~Acernese$^\text{19a,19c}$}\noaffiliation\author{C.~Adams$^\text{31}$}\noaffiliation\author{R.~Adhikari$^\text{29}$}\noaffiliation\author{P.~Ajith$^\text{29}$}\noaffiliation\author{B.~Allen$^\text{2,79}$}\noaffiliation
% ...

\collaboration{$^\ast$The LIGO Scientific Collaboration and $^\dagger$The Virgo Collaboration}
\noaffiliation




\maketitle


\section{Introduction}\label{sec:introduction}
This paper presents a search for gravitational waves from \ac{BBH} coalescences
with total mass $25\,\Msun\le M\le 100\,\Msun$ and component masses
$1\,\Msun\le m_1,m_2\le 99\,\Msun$. The search used 
\ac{LIGO}~\cite{Abbott:2007kv} data taken during
the \ac{S5} from November, 2005 to September 2007 when both \ac{LIGO} sites
were operating.  The first \ac{LIGO} site in Hanford, Washington hosts two
interferometric gravitational-wave detectors, a 4km detector, H1, and a 2km
detector, H2.  The second site in Livingston, Louisiana hosts a single 4 km
detector, L1.  The Virgo detector~\cite{Acernese:2008b} 
in Cascina, Italy commenced its first science
run (VSR1) on May 18, 2007 and since then \ac{LIGO} and Virgo have operated
their instruments as a network.  However, this search did not use Virgo data,
because it was not as sensitive to these high mass systems during VSR1.
Additionally the GEO600 detector in Germany was functioning during \ac{S5}.
However, GEO600 data was not analyzed for similar reasons.  The search
results for compact binaries with total mass $M \le 35\,\Msun$ in \ac{LIGO}
\ac{S5} and Virgo VSR1 data have been reported
in~\cite{Collaboration:2009tt,Abbott:2009qj, S5LowMassLV}.  To date no
gravitational waves have been directly observed.  

The gravitational-wave driven evolution of \ac{BBH}s is conventionally split
into three stages -- \ac{IMR}.  The gravitational-wave signal during the
adiabatic inspiral phase can be described by \ac{PN} expansion.  This technique
is very accurate for comparable-mass systems at large separations, but breaks
down near the \ac{ISCO}\footnote{The \ac{ISCO} occurs at a gravitational-wave
frequency of $f_{\text{ISCO}}\approx 4.4\times
10^3\un{Hz}\times(M/\Msun)^{-1}$ for a test particle orbiting a Schwarzschild
black hole with mass $M$. It is typically taken to define the onset of the
merger epoch.}. Modeling of the merger requires the numerical solution of the
full Einstein equations in a highly dynamical strong-field regime.  After the
merger, the rapidly damped quasi-normal ringdown of the \ac{BH} toward a
stationary Kerr black hole is described by black-hole perturbation theory.
This is the first analysis that incorporates a template family of waveforms
modeling all three stages of \ac{BBH} coalescence.  This search covers systems
for which the effects of \ac{BH} spins can be neglected for detection.  For
\ac{BH}s in this mass range the merger occurs in the LIGO detectors' most
sensitive frequency band.  

%
%

\subsection{Motivation to search for higher mass systems}
Black holes observed in X-ray binaries range up to $\sim 20\,\Msun$
\cite{Remillard:2006fc, Orosz:2007, Ozel:2010, Farr:2010}, and predictions from population-synthesis
models have typically suggested the masses of components of BH-BH binaries that
merge within $10\un{Gyr}$ will mostly lie in the range $5\,\Msun\lesssim
m_1,m_2 \lesssim\,10\Msun$ \cite{BulikBelczynski:2003, OShaughnessy:2009}.
However, a number of channels have been suggested through which significantly
more massive black-hole binaries could form.  

Observations of IC10 X-1, a binary with a massive black-hole ($\gtrsim 24\
\Msun$) accreting from a Wolf-Rayet companion star, and a similar
recently-observed binary NGC 300 X-1 \cite{Crowther:2010}, suggest that more
massive BH-BH binaries can form through isolated binary evolution, with
component masses $\sim 20\,\Msun$ \cite{Bulik:2008}.  Meanwhile, several
simulations over the past few years have indicated that dynamical formation
could significantly contribute to coalescence rates involving BH-BH binaries in
dense stellar environments, such as globular and nuclear star clusters
\cite{OLeary:2007, Sadowski, MillerLauburg:2008, OLeary:2008}.  The most
massive black holes are likely to sink to the centers of clusters through mass
segregation and substitute into binaries during three-body encounters, thus
favoring relatively massive components in dynamically formed BH-BH binaries.
Moreover, the BH merger products in such dense clusters can be reused if they
are not ejected from the cluster due to recoil kicks, leading to higher-mass
mergers in subsequent generations; component masses for BH-BH mergers in
globular clusters can therefore range to $\sim 30\,\Msun$ \cite{Sadowski} and
beyond.  Additionally, although stellar winds in high-metallicity environments
may prevent the formation of massive black holes, mass loss through stellar
winds would be much less significant in low-metallicity environments, allowing
more massive black holes to form \cite{Belczynski:2009, Belczynski:2010}.

Binaries including an intermediate-mass black hole (IMBH) having a mass
$50\,\Msun\lesssim m\lesssim 500\,\Msun$ could represent another exciting
source for LIGO and Virgo detectors.  Observational evidence for IMBHs is still
under debate (see reviews \cite{2004IJMPD..13....1M, miller-2008} for
additional details), although a recently discovered ultra-luminous X-ray source
\cite{2009Natur.460...73F} represents a possible IMBH detection.  If IMBHs do
exist in the centers of some globular clusters, they could contribute to
coalescence rates in one of three ways: (i) through inspirals of stellar-mass
compact objects into IMBHs \cite{Brown:2007, Mandel:2007rates}; (ii) through
mergers of two IMBHs that formed in the same stellar cluster
\cite{imbhlisa-2006}; and (iii) through mergers of IMBHs from two different
globular clusters when their host clusters merge \cite{Amaro:2006imbh}.  It may also be possible to detect mergers of binary IMBHs arising from a direct collapse of early population III binary stars \cite{Belczynski:2004popIII}.  Although the rates of events involving IMBHs are highly uncertain, they may reach tens of detections per year in the Advanced LIGO/Virgo era (see \cite{ratesdoc} for a
review of expected detection rates for all binary types relevant to the
advanced-detector era).

Advances in \ac{NR} in the last 5 years have enabled this search for \ac{BBH}
coalescence to significantly improve on previous methods and results.  We
describe those advances and the impact on gravitational-wave astronomy in the
next section.

%
%

\subsection{Complete inspiral, merger and ringdown waveforms from 
numerical relativity}
\label{sec:BreakthroughNR}
Constructing nonperturbative numerical solutions for the merger of two black
holes has proven to be remarkably difficult. It has taken more than four
decades since Hahn and Lindquist first attempted the numerical investigation
of colliding black holes~\cite{Hahn64} to compute the gravitational-wave
signal from the last orbits, merger and ringdown of a black-hole binary
system.  These simulations are now possible using many different methods.
Only months after Pretorius' initial breakthrough in
2005~\cite{Pretorius:2005gq}, his success was repeated using a different
approach~\cite{Campanelli:2005dd, Baker:2005vv}, and since then several
\ac{NR} groups were able to produce increasingly accurate \ac{BBH} simulations
exploring increasing portions of the parameter space (see,
e.g.~\cite{Pretorius:2007nq, Husa:2007zz, Hannam:2009rd,Hinder:2010vn} for
recent overviews on the field, and Sec. 2 of~\cite{Aylott:2009ya} for a
compact summary).  The success of \ac{NR} simulations has lead to a range of
new physical insights including the calculation of recoil velocities produced
by asymmetric emission of gravitational radiation during the merger
process~\cite{Herrmann:2006ks, Baker:2006vn, Gonzalez:2006md, Herrmann:2007ac,
Koppitz:2007ev, Campanelli:2007ew, Gonzalez:2007hi, Tichy:2007hk,
Campanelli:2007cga, Baker:2007gi, Herrmann:2007ex, Brugmann:2007zj,
Schnittman:2007ij, Pollney:2007ss, Lousto:2007db, Baker:2008md, Dain:2008ck,
Healy:2008js, Gonzalez:2008bi} and the prediction of the parameters of the
remnant Kerr \ac{BH} for a wide class of initial
configurations~\cite{Campanelli:2006fg, Gonzalez:2006md, Campanelli:2006fy,
Berti:2007fi, Rezzolla:2007xa, Boyle:2007sz, Rezzolla:2007rd,
Marronetti:2007wz, Rezzolla:2007rd, Sperhake:2007gu, Hinder:2007qu,
Berti:2007nw, Boyle:2007ru, Tichy:2008du, Rezzolla:2008sd}.  Most importantly
for the \ac{GW} community, these simulations were able to compute accurate
\ac{GW} waveforms from the late inspiral and merger of \ac{BBH}s in many
configurations, and these predictions were successfully matched to \ac{PN} and
\ac{EOB} predictions~\cite{Baker:2006ha, Hannam:2007ik, Boyle:2007ft,
Hannam:2007wf, Campanelli:2008nk, Boyle:2008ge,hinder-2008,chu-2009,
pollney-2009,Buonanno:2006ui,Pan2007,Buonanno:2007pf,DN2007b,DN2008,Boyle2008a,
Damour2009a,Buonanno:2009qa,Ajith:2007qp,2009arXiv0909.2867A}.
Thus, by combining analytical and numerical calculations, it is now possible
to construct accurate waveform templates coherently modeling the \ac{IMR} of
the coalescence of \ac{BBHs}, as described in section \ref{sec:templates}.  

%
%

\subsection{Summary of past searches}
The \ac{LIGO} and Virgo Scientific collaborations previously searched for
systems that are a subset of the \ac{BBH} parameter space explored by this
analysis using different techniques.   Previous searches used templates that
modeled only the inspiral or ringdown phases.  None employed templates that
include \ac{IMR} waveforms.  To date no gravitational waves from \ac{BBH}
coalescence have been observed.  

The first search explored systems with component masses in the range $3\,\Msun
\le m_1,m_2\le 20\,\Msun$ in $\sim 386$ hours of \ac{LIGO}'s second science
(S2) run~\cite{LIGOS2bbh} with a 90\% sensitivity to systems up to 1\,Mpc.  The
second search covered sources up to 40 $\Msun$ total mass for \ac{LIGO}'s third
science run (S3, 788 hours) and up to 80 $\Msun$ in total mass for the fourth
science run (S4, 576 hours)~\cite{LIGOS3S4all} with $\sim 10$ times the range
sensitivity of S2, and placed a 90\% confidence upper limit on the merger rate
of $\sim 0.3\,L_{10}^{-1}\un{yr}^{-1}$ for systems with a total mass of $\sim
40\,\Msun$.  (Here $L_{10}$ is $10^{10}$ times the blue Solar luminosity and is
used as a proxy for the expected number of sources in a galaxy.  For searches
that extend beyond $\sim20\un{Mpc}$ there are approximately
$0.0198\,L_{10}\un{Mpc}^{-3}$\cite{LIGOS3S4Galaxies}.  The presently discussed
search extends beyond $20\un{Mpc}$ and has sufficient sensitivity to use units
of $\text{Mpc}^{-3}\un{Myr}^{-1}$.  Additionally the Milky Way is
$\sim1.7\,L_{10}$.) The S2, S3 and S4 science runs used phenomenological
waveforms proposed by~\cite{BuonannoChenVallisneri:2003a} that extended the
inspiral to higher frequency but did not include the complete \ac{IMR} 
signal nor the
effects of spin.  A search that used templates including the spin effects was
conducted over S3 data targeting asymmetric systems with component masses
$1\,\Msun\le m_1\le 3\,\Msun$ and $12\,\Msun\le m_2\le 20\,\Msun$
\cite{S3_BCVSpin}.  

\ac{BBH} mergers with sufficiently high mass will have most of their in-band
gravitational-wave amplitude in the ringdown phase.  A search over S4 data
probed the ringdown phase of \ac{BBH} coalescence and placed  90\% confidence
upper limits on the merger rate for systems with total masses $85\,\Msun\le
M\le 390\,\Msun$ of $1.6 \times 10^{-3}\,L_{10}^{-1}\un{yr}^{-1}$ ($32
\un{Mpc}^{-3} \un{Myr}^{-1}$)~\cite{Abbott:2009km}.  The ringdown waveforms
are a function of the final state of the black hole and do not depend on the
details of the merger.  For this reason a spectrum of initial states (arbitrary
spins and component masses) can be probed via a ringdown-only search. 

Finally, the \ac{LSC} and Virgo searched S5 data and Virgo's first science run
data (VSR1, which overlapped with the last $\sim$ 6 months of S5) for \ac{BBH}s
with a total mass up to 35
M$_{\odot}$~\cite{Collaboration:2009tt,Abbott:2009qj, S5LowMassLV}.  The 90\%
confidence upper limit on merger rate for black hole binaries with total mass
$\sim 30$ $M_{\odot}$ was $\sim 3 \times 10^{-4} \,L_{10}^{-1}\un{yr}^{-1}$,
which is $\sim6$ Mpc$^{-3}$ Myr$^{-1}$~\cite{S5LowMassLV}.

%
%

\subsection{Summary of the present results}
No plausible gravitational-wave signals were detected in this search.  The
loudest events are discussed in section \ref{sec:candidates}.  Despite not
detecting \ac{BBH} signals directly we are able to infer an upper limit on the
merger rate of such systems in the nearby Universe.  We do not impose a
particular population model within our mass range and instead present our
merger rate limits as a function of component mass ranges.  To 90\% confidence
we constrain the rate of mergers for \ulrange binary black hole systems to be
no more than \ulnum.  We highlight numbers from this range because it may include some of the heavier BH binaries that may arise in population synthesis models (e.g., \cite{Belczynski:2010}), but  
was not covered by the S5 low-mass search \cite{Collaboration:2009tt,Abbott:2009qj, S5LowMassLV}.
Additional mass pair rate limits are given in section \ref{sec:upper limits}.

The paper organization is as follows. In section \ref{sec:templates} we
describe two families of waveforms used in this search (effective one-body
model and phenomenological IMR model).  Section \ref{sec:pipeline} summarizes
the key points of our data analysis pipeline. This part includes information
about the template bank, data quality, background estimation and candidate
ranking statistics.  Section \ref{sec:candidates} contains results of the
search, in particular we present the loudest events. In section \ref{sec:upper
limits} we discuss detection efficiency, which we estimate by injecting
simulated signals into the detector data.  Additionally in this section, we
present an upper limit on the coalescence rate for this search.  Finally,
section \ref{sec:conclusions} presents the conclusions and plans for future
improvements.


\section{Waveforms used in this search}\label{sec:templates}

Modeled waveforms are invaluable tools for extracting weak signals from noisy
data in gravitational-wave searches for compact binaries~\cite{Cutler:1992tc}.
The models are used to efficiently filter the data for signals and to assess
the sensitivity of the instruments and data analysis procedure via simulations.
This section motivates the need for new waveform models in this search and
describes the models chosen.

\begin{figure}
\includegraphics[width=0.5\textwidth]{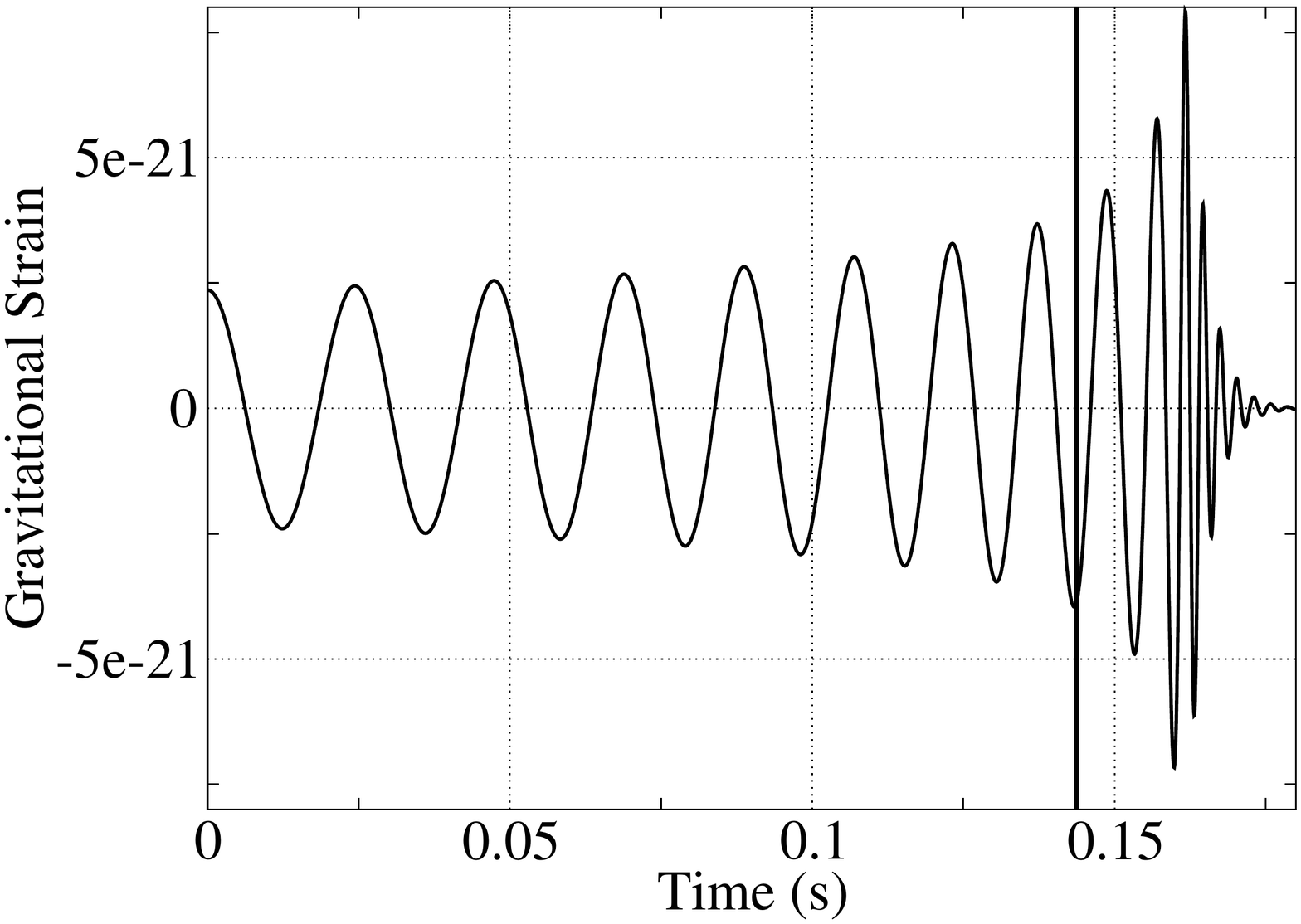}
\includegraphics[width=0.5\textwidth]{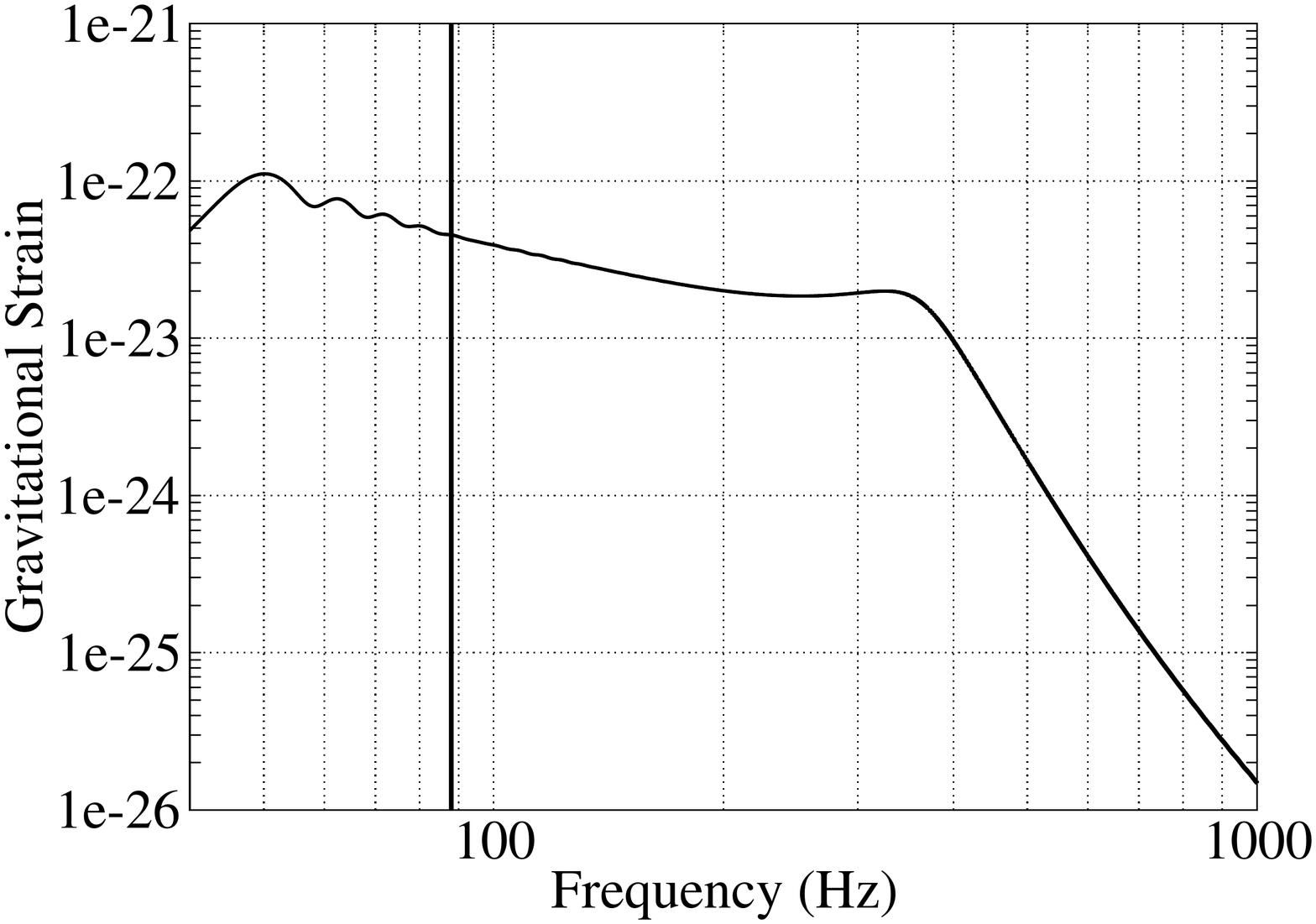}
\caption{\label{fig:waveform}Example  of the \emph{EOBNRv1} IMR waveforms 
used in this search for a $(25+25)\,M_\odot$ binary optimally located 
and oriented at $100$ Mpc in the time domain (top panel) and the 
frequency domain (bottom panel). The solid vertical lines mark the location of 
the Schwarzschild ISCO, which is the termination point 
for inspiral-only waveforms.}
\end{figure}

For binaries with a total mass in the range targeted by this search, 
$25 - 100\, \Msun$, the \ac{ISCO} is reached in the sensitive frequency 
band of the LIGO interferometers. Thus, standard inspiral-only \acs{PN} 
waveforms, which are typically terminated at the \ac{ISCO} frequency, do not 
capture all of the observable signal. Furthermore, this abrupt, in-band 
end of the search templates can be problematic for the signal consistency
checks. On the 
other hand, \ac{IMR} templates model all of the observable signal and naturally 
decay away during the ringdown phase rather than abruptly ending. For these 
reasons, it is highly desirable to use \ac{IMR} template waveforms to search 
for binary coalescences in this mass range. In figure~\ref{fig:waveform}, 
we plot an example IMR waveform in the time and frequency domains 
and note the extra signal relative to inspiral-only waveforms.

Fortunately, the recent breakthroughs in numerical relativity (see
Sec.~\ref{sec:BreakthroughNR}) have revealed the nature of the merger and
ringdown phases of BBH coalescences. While it is infeasible to use the NR
simulations directly as search templates, insights gained from the simulations
have informed the development of analytic \ac{IMR} waveform models. Currently,
two main paradigms exist in the construction of IMR waveforms. In the \ac{EOB}
approach, an effective-one-body description of the two-body problem is tuned
with NR simulations and then matched to the quasi-normal modes of the BH
ringdowns to produce analytical IMR waveforms in the time domain.  In the
phenomenological IMR model, the NR waveforms are matched to PN waveforms to
produce ``hybrid'' PN-NR waveforms which are then parametrized to produce
analytical IMR templates in the frequency domain. The \ac{EOB}
waveforms are used as
search templates and also as injected waveforms to test our detection
efficiency. The phenomenological waveforms are used for injections and provide
a check that our search pipeline can detect waveforms which are slightly
different than our search templates. The next two subsections describe each of
these families of analytic \ac{IMR} waveforms. 

Since \ac{EOB} waveforms are generated in the time domain, the presence of an
abrupt starting point at a given low frequency can result spurious high
frequency power.  To mitigate these effects a tapering window was applied to
the beginning of the generated \ac{EOB} waveforms~\cite{MRS:2010}.

\subsection{Effective one-body model}\label{sec:EOB}

The \ac{EOB} approach, originally introduced
in~\cite{BuonannoDamour:1999,BuonannoDamour:2000}, provides a
PN-resummed Hamiltonian which can be used to evolve a binary system
through its inspiral and the final ``plunge'' of the compact objects
before they merge.  This trajectory can be used to generate a waveform
$h^{\text{insp-plunge}}(t)$ which can be matched onto a waveform
$h^{\text{merger-RD}}(t)$ describing the merger and ringdown of the
resulting black hole, made up of a superposition of the black hole's
quasi-normal modes.  The two pieces are combined at a suitably chosen
matching time $t_{\text{match}}$ to produce an
inspiral-plunge--merger-ringdown EOB
waveform~\cite{BuonannoDamour:2000}
\begin{equation}
\label{eobfullwave}
h(t) = h^{\text{insp-plunge}}(t)\,
\Theta(t_{\text{match}} - t) + 
h^{\text{merger-RD}}\,\Theta(t-t_{\text{match}})\,,
\end{equation}
where $\Theta()$ is the Heaviside step function.

The inspiral-plunge \ac{EOB} waveform at leading-order amplitude in a PN
expansion is determined from the trajectory $r(t)$, $\phi(t)$
as~\cite{BuonannoDamour:2000}
\begin{equation}\label{inspwave}
{h}^{\text{insp-plunge}}(t) \equiv \frac{4 G M \eta}{D_L c^2}\, \left(\frac{G M}{c^3}\frac{d\phi}{dt}\right)^{2/3}
\,\cos [2 \phi(t)]\,,
\end{equation}
where $D_L$ is the luminosity distance. 
We now summarize the fundamentals of the \ac{EOB} calculation of the
trajectory; more details can be found in
~\cite{BuonannoDamour:1999,BuonannoDamour:2000,DamourJaranowskiSchaefer:2000,Damour03,Buonanno:2006ui,Pan2007,Buonanno:2007pf,DN2007b,DN2008,Boyle2008a,Damour2009a,Buonanno:2009qa}.
As usual, $m_1$ and $m_2$ are the black hole masses, $M=m_1+m_2$ is
the total mass of the binary, $\mu = m_1\,m_2/M$ is the reduced mass,
and $\eta = \mu/M$ is the symmetric mass ratio.

For a binary with negligible spin effects, the motion is confined to a
plane and can be described in the center of mass by polar coordinates
$(r,\phi)$.  The conservative dynamics is then captured by a
Hamiltonian $H^{\text{EOB}}(r,p_r,p_\phi)$.  The trajectory is evolved
according to Hamilton's equations~\cite{BuonannoDamour:2000}
\begin{subequations}
  \label{eq:eobhami}
\begin{align}
  \frac{dr}{dt} &= \frac{\partial H^{\text{EOB}}}
  {\partial p_{r}}(r,p_{r},p_\phi)\,, 
  \label{eq:eobhamone} \\
  \frac{d \phi}{dt}  &=
  \frac{\partial H^{\text{EOB}}}
  {\partial p_\phi}(r,p_{r},p_\phi)\,, \\
  \frac{d p_{r}}{dt} &= - \frac{\partial H^{\text{EOB}}}
  {\partial r}(r,p_{r},p_\phi) \,, \\
  \frac{d p_\phi}{dt} &=
  \mathcal{F}_\phi(r,p_{r},p_{\phi})\,,
  \label{eq:eobhamfour}
\end{align}
\end{subequations}
The inspiral of the binary comes about due to the addition of
non-conservative dynamics in the last of Hamilton's equations via the tangential
radiation-reaction force $\mathcal{F}_\phi$ arising from the basic \ac{PN}
expression of the energy flux. Here we use a Keplerian Pad\'e resummation~\cite{Damour:1998zb} 
of the energy flux as given by Eq.~(15) of~\cite{Buonanno:2007pf}.
More recent models have used more sophisticated fluxes, such as the 
$\rho$-resummation~\cite{DIN} and non-Keplerian flux models which describe 
non-quasi-circular effects~\cite{DN2007b,DN2008,Damour2009a,Buonanno:2009qa}.

The form of the \ac{EOB} (resummed) Hamiltonian is~\cite{BuonannoDamour:1999} 
\begin{equation}
\label{himpr}
H^{\text{EOB}}(r,p_{r},p_\phi)
= Mc^2\,\sqrt{1 + 2\eta\,\left ( \frac{H^{\text{eff}}}{\mu c^2} - 1\right )} \,,
\end{equation}
where $H^{\text{eff}}$ is the effective 
Hamiltonian~\cite{BuonannoDamour:1999,DamourJaranowskiSchaefer:2000} 
\begin{equation}
  \label{eq:genexp}
  \begin{split}
  H^{\text{eff}}
  = \mu c^2
  \Biggl(
    A(r)
    \Biggl[&
      1 + \frac{A(r)}{D(r)} \frac{p_r^2}{M^2c^2}
      + \frac{p_\phi^2}{M^2c^2r^2} 
      \\
      &+ 2(4-3\eta)\eta\frac{G^2 p_{r}^4}{M^2 c^8 r^2}
    \Biggr]
  \Biggr)^{1/2}
  \,.    
  \end{split}
\end{equation}
and where the radial potential functions $A(r)$ and $D(r)$ appear in the 
effective metric~\cite{BuonannoDamour:1999} 
\begin{equation}
  ds_{\text{eff}}^2 =
  -A(r)\,c^2\,dt^2 + \frac{D(r)}{A(r)}\,dr^2 +
  r^2\,\Big(d\theta^2+\sin^2\theta\,d\phi^2\Big) \,.  
\label{eq:EOBmetric}
\end{equation}
The Taylor-approximants to the coefficients $A(r)$ and $D(r)$ can be
written as
\begin{subequations}
\begin{align}
  \label{coeffA}
  A_{k}(r) &= \sum_{i=0}^{k+1} a_i(\eta) \left(\frac{GM}{rc^2}\right)^i\,,\\
  \label{coeffD}
  D_{k}(r) &= \sum_{i=0}^k  d_i(\eta) \left(\frac{GM}{rc^2}\right)^i\,.
\end{align}
\end{subequations}
The functions $A(r)$, $D(r)$, $A_k(r)$ and $D_k(r)$ all depend on the
symmetric mass ratio $\eta$ through the $\eta$--dependent coefficients
$a_i(\eta)$ and $d_i(\eta)$. [When $\eta \rightarrow 0$,
$A(r)\rightarrow 1-\frac{2 GM}{rc^2}$ and $D(r)\rightarrow 1$ and the metric
(\ref{eq:EOBmetric}) reduces to the Schwarzschild metric.] These
coefficients are currently known through 3PN order (i.e., up to $k=3$)
and can be found in ~\cite{Buonanno:2007pf}. During the last stages of
inspiral and plunge, the \ac{EOB} dynamics can be adjusted closer to the
numerical simulations by including in the radial potential $A(r)$ a
pseudo 4PN coefficient $a_5(\eta)=a_5\,\eta$, with $a_5$ a
constant. Here, we follow ~\cite{Buonanno:2007pf} and fix $a_5 = 60$.
We refer to this model, the first \ac{NR}-adjusted \ac{EOB} model
implemented for a search of \ac{GW} data, as \emph{EOBNRv1}.
Since ~\cite{Buonanno:2007pf} was published, more accurate numerical
simulations became available and more sophisticated \ac{EOB} models
have been calibrated. This includes a different value of
$a_5$~\cite{DN2007b,DN2008,Boyle2008a,Buonanno:2009qa} and also the
introduction of a pseudo 5PN coefficient
$a_6(\eta)=a_6\,\eta$~\cite{Damour2009a}, with $a_6$ a constant.  We refer 
to the second \ac{NR}-adjusted \ac{EOB} model implemented for a search of \ac{GW} data,
as \emph{EOBNRv2}. This most recent EOB template family has been 
developed in \cite{Pan:2011};  it includes    
the latest improvements~\cite{DN2007b,DN2008,Boyle2008a,Buonanno:2009qa} to the EOB model and also other refinements which are necessary to match highly-accurate 
NR waveforms for a broad range of mass ratios.

In order to assure the presence of a horizon in the effective metric
(\ref{eq:EOBmetric}), a zero needs to be factored out from $A(r)$. This is
obtained by applying a Pad\'e resummation~\cite{DamourJaranowskiSchaefer:2000}.
The Pad\'e resummations of $A(r)$ and $D(r)$ at pseudo 4PN
order are denoted  $A_4^1(r)$ and $D_4^0(r)$
\footnote{We note that Ref.~\cite{Buonanno:2007pf} contains a typo 
and should read\\ 
$D_4^0(r) = \frac{r^4}{r^4 + 6\eta r^2 + 2 \eta (26 - 3 \eta) r + 36 \eta^2}$}, and the explicit form used in
this paper can be read from ~\cite{Buonanno:2007pf}.

The merger-ringdown waveform in the EOB approach is built as a superposition of
quasi-normal modes, ~\cite{BuonannoDamour:2000,Damour06,Buonanno:2006ui}
\begin{equation}
\label{RD}
h^{\text{merger-RD}}(t) = \sum_{n=0}^{N-1} A_n\,
e^{-i\sigma_n (t-t_{\text{match}})},
\end{equation}
where $n$ is the overtone number of the Kerr quasi-normal mode, $N$ is the
number of overtones included in our model, and $A_n$ are complex amplitudes to
be determined by a matching procedure described below. We define $\sigma_{n}
\equiv \omega_{n} - i \alpha_{n}$, where the oscillation frequencies
$\omega_{n}>0$ and the inverse decay-times $\alpha_{n}>0$, are numbers
associated with each quasi-normal mode. The complex frequencies are known
functions, uniquely determined by the final black-hole mass and spin. They can
be found in ~\cite{2006PhRvD..73f4030B}. The final black-hole masses and spins
are obtained from the fitting to numerical results worked out in
~\cite{Buonanno:2007pf}.

The complex amplitudes $A_{n}$ in Eq.~(\ref{RD}) are determined by matching the
EOB merger-ringdown waveform with the EOB inspiral-plunge waveform close to the
EOB light ring. In particular, here we use the matching point which is provided
analytically by Eq.~(37) of ~\cite{Buonanno:2007pf}. In order to do this, we
need $N$ independent complex equations that are obtained at the matching time
by imposing continuity of the waveform and its time derivatives,
\begin{equation}
\frac{d^k}{dt^k}h^{\text{insp-plunge}}(t_{\text{match}})=
\frac{d^k}{dt^k}h^{\text{merger-RD}}(t_{\text{match}})\,, 
\end{equation}
with $k=0,1,2,\cdots, N-1$. In this paper we use N=3.  The above matching
approach is referred to as {\it point matching}. Although it gives better
smoothness around the matching time, it is not very stable numerically when $N$
is large and higher-order numerical derivatives are employed. More
sophisticated matching procedures have been proposed in the literature to
overcome the stability issue~\cite{DN2007b,DN2008,Damour2009a,Buonanno:2009qa}, and 
will be adopted in the future. 

\subsection{Phenomenological IMR model}\label{sec:phenom}
Reference ~\cite{Ajith:2007kx} presented a different way of constructing
non-spinning IMR waveforms by combining PN calculations with numerical
simulations. They first constructed a family of hybrid waveforms by matching PN
waveforms with NR waveforms in certain overlapping time intervals where both
the approaches are expected to be valid~\cite{Ajith:2007qp}.  Restricted 3.5PN
waveforms in the TaylorT1 approximation were matched to NR waveforms produced
by the \texttt{BAM} NR code~\cite{Bruegmann:2008}. These hybrid waveforms were
used to construct a family of analytical waveforms in the Fourier domain, of
the form ${\tilde h}(f) \equiv {A}_{\text{eff}}(f) \, e^{i\Psi_{\text{eff}}(f)},$ where the effective amplitude and phase are expressed as:
\begin{subequations}
\begin{align}
{A_{\text{eff}}}(f) \equiv&\ C
\begin{cases}
\left(f/f_{\text{merg}}\right)^{-7/6}   & f < f_{\text{merg}}\\
\left(f/f_{\text{merg}}\right)^{-2/3}   & f_{\text{merg}} \leq f < f_{\text{ring}}\\
w \, {\cal L}(f,f_{\text{ring}},\sigma) & f_{\text{ring}} \leq f < f_{\text{cut}}
\end{cases}
\\
\begin{split}
\Psi_{\text{eff}}(f) \equiv&\ \frac{1}{\eta}\,\sum_{k=0}^{7} 
(x_k\,\eta^2 + y_k\,\eta + z_k) \,(\pi M f)^{(k-5)/3} \\ 
&+ 2 \pi f t_0 + \varphi_0\,.
\end{split}
\end{align}
\label{eq:phenWaveAmpAndPhase}
\end{subequations}
In the above expressions, $C$ is a numerical constant whose value depends on
the location and orientation of the binary as well as the physical parameters,
${\cal L}(f,f_{\text{ring}},\sigma)$ is a Lorentzian function that has a  width
$\sigma$, and that is centered around the frequency $f_{\text{ring}}$.  The
normalization constant $w$ is chosen so as to make ${A}_{\text{eff}}(f)$
continuous across the ``transition'' frequency $f_{\text{ring}}$. The parameter
$f_{\text{merg}}$ is the frequency at which the power-law changes from $f^{-7/6}$
to $f^{-2/3}$.  The phenomenological parameters $\mu_j \equiv \{
f_{\text{merg}}, f_{\text{ring}}, \sigma, f_{\text{cut}} \}$ are given in terms
of the physical parameters of the binary as:  $\pi M \mu_j =  a_j \, \eta^2 +
b_j \, \eta + c_j$. The coefficients $\{a_j, b_j, c_j|j=0\ldots3\}$ and
$\{x_k,y_k,z_k|k=0,2,3,4,6,7\}$ are tabulated in Table I of ~\cite{Ajith:2007xh}.  We refer to the waveform family defined by these coefficients as \emph{IMRPhenomA}, and these are the waveforms used for injections in the present search.  These waveforms are generated in the frequency domain and are then converted to the time domain for injections by means of the inverse Fourier transform.  

The choice of the time interval for matching PN and NR waveforms is somewhat
\emph{ad hoc}.  Currently, the matching interval is chosen so as to maximize
the fit of PN and NR waveforms.  Moreover, the PN waveforms employed in the
matching are computed in the \emph{restricted} PN approximation, and the
amplitude of the NR waveforms is scaled to match with PN waveforms. This causes
the amplitude of the waveforms to have a systematic bias of $\sim10\%$. Later improvements in this model have already addressed some of these issues~\cite{2009arXiv0909.2867A}; we refer to this improved waveform family as \emph{IMRPhenomB}. 

\subsection{Systematic errors in waveform models }\label{sec:modelerrors}

Although the two waveform families have been tested via comparisons to
numerical waveforms, there are a number of possible sources of systematic
uncertainty in the two waveform families.  For example, as discussed above,
there are subtleties in choosing the matching interval between PN inspiral
waveforms and numerical simulations when constructing the hybrid waveforms
used to calibrate the Phenomenological waveforms. Similarly, the EOB 
procedure to attach the merger-ringdown waveform to the inspiral-plunge
waveform can be quite delicate and become unstable if not done properly. 
Most notably, both waveform families have been tested against numerical
simulations only in the nearly-equal-mass regime, up to ratios of $3:1$ and
$4:1$.  It is not clear whether these waveforms are faithful to the actual
signals in the case of highly unequal masses.  

The waveform models initially used for this search, \emph{EOBNRv1} and
\emph{IMRPhenomA}, had both been revised by the time the search neared
completion.  The revisions, which included improved calibration and more
accurate matching to NR waveforms, as well as improved modeling of the
post-Newtonian inspiral phase, resulted in waveforms that were significantly
more faithful.  For example, the revised version of the phenomenological
waveform family, \emph{IMRPhenomB}, has systematic biases in \ac{SNR} of  $\lesssim 10\%$
relative to NR waveforms in the mass range of interest and for mass ratios
below $4:1$; for the revised version of the EOB waveform family,
\emph{EOBNRv2}, the systematic biases in \ac{SNR} relative to NR waveforms are $\lesssim
3\%$.  These systematic biases do not account for any errors in the NR
waveforms themselves.  

The largest effect of the revision of both models was to
systematically reduce the gravitational-wave amplitude during merger.
We found that within our errors it was sufficient to adjust the
distance of the simulated signals to take into account the lower
intrinsic gravitational-wave amplitude in the corrected models.  The
upper limits quoted in section \ref{sec:upper limits} are thus based
on a search carried out with \emph{EOBNRv1} templates, but with the
distances of \emph{EOBNRv1} and \emph{IMRPhenomA} injections adjusted
to match the \ac{SNR} of the revised \emph{EOBNRv2} and \emph{IMRPhenomB}
waveform models.

We can get a sense of the systematic uncertainty in the waveform amplitudes by
comparing the SNRs between the two waveform families.   We find that the SNR of
the most recent versions of the two families, \emph{EOBNRv2} and
\emph{IMRPhenomB}, agrees to better than $\sim 10\%$ for mass ratios less than
$6:1$ in the mass range of interest, but diverges by nearly $50\%$ for mass
ratios of $10:1$.  The latter value is chosen as the limit on the mass ratio
for phenomenological IMR injections.


\section{The Data Analysis Procedure}\label{sec:pipeline}

The data analysis procedure involves a multi-stage pipeline that automates the
extraction of signals from the data, the analysis of coincident events between
detectors and the estimation of background.  The pipeline used for this search
was similar to that of previous \ac{S5} searches
\cite{Collaboration:2009tt,Abbott:2009qj, S5LowMassLV} except for the choice of
template waveforms and some minor points described in subsequent sections 

Here we summarize the data analysis procedure. First, data for the three
different detectors, H1, H2, and L1 are divided into 2048 s blocks in order to
estimate the time dependent power spectral density (PSD) of the detector noise.
The PSD is required to choose the search templates and to filter the data
itself.  Next, the data are processed in a two-stage procedure.  The first
stage filters the data with the templates and identifies potential events in
each detector.  Then the pipeline checks for coincidence between detectors. We
allow double and triple coincident combinations between detectors.   After
finding coincident events the data are re-filtered using only the templates
that participated in the coincident events.  The data needs to be filtered with
fewer templates  at the second stage after demanding coincidence with other
detectors.  The second filtering stage employs the $\chi^2$
veto~\cite{Allen:2004}, which drastically reduces the background of this
search, but is too computationally expensive to be performed during the first
filtering stage with the full template bank.  Once coincident events are
identified they are clustered in a ten second window to produce a maximum of
one coincident event every ten seconds.  We apply the same procedure to
time-shifted data streams and compare the time-shifted results to the zero-lag
results to assess the significance of our events.  This procedure is repeated
with simulated signals in order to assess the sensitivity of the pipeline. 

In the remaining sections, we elaborate on this procedure emphasizing
differences with previously published searches.

\subsection{Generation of coincident event candidates}

In this section we describe the process of obtaining candidate events.  First
we discuss how to choose templates to filter the data.  Next we describe the
filter process itself and how to identify events that are significant in a
single detector.  We then describe how we check for coincident events between
detectors.  Finally we describe how data quality impacts our assessment of
candidates. 

\subsubsection{Selection of search template parameters}

The observed gravitational waveform depends on the component masses of the
binary.  A bank of template waveforms called a template bank is chosen to
adequately cover the parameter space of possible waveforms. The template bank
used for the search consisted of templates covering total mass between 25--100
{\Msun}, and component masses between 1--99 {\Msun}.  The bank was tiled using a
hexagonal placement algorithm~\cite{hexabank}, such that the intended minimum
\ac{SNR} was 97\% of its maximal value\cite{Owen:1995tm}.  The template spacing
was determined using the metric calculated for the stationary phase
approximation~\cite{BBCCS:2006} extended to the effective ringdown frequency.
This metric, terminated at \ac{ISCO} frequency, was used in previous searches
for signals from low-mass systems.  Although the metric is not formally correct
for the \ac{EOB} templates used in this search it has been found that the bank
provides the desired minimal match for most of the parameter space and at worst
a 95\% match for the high-mass region of the bank.  The average number
of templates required to cover this parameter space was $\sim1600$ per detector.

\subsubsection{Filtering}

After properly selecting the templates to cover the mass parameter space, the
data are filtered.  The signal to noise ratio for a given template waveform
$h(t)$ is a convolution of the template with the data weighted by the noise
power spectral density, defined as
%
%
\begin{subequations}
\begin{align}
\label{eq:filter}
z(t) &= 4 
             \int_0^\infty \frac{\tilde{h}(f)^* \tilde{s}(f)}
                      {S_n(f)} \, e^{2\pi i f t} \, d\!f ,\\
\sigma^2 &= 4 \int_0^\infty \frac{\tilde{h}(f)^* \tilde{h}(f)}
                      {S_n(f)} \, d\!f ,
\end{align}  
\end{subequations}
where the tilde and $^*$ denote a Fourier transform and a complex conjugate,
$s$ refers to the data and $S_n$ is the noise power spectral density. In order
to cover the entire parameter space all of the data are filtered with every
template.  $z(t)$ is a complex time series where the real part represents a
template phase of 0 and the imaginary part represents a phase of $\pi/2$.  The
real-valued \ac{SNR} $\rho$ is given by $\rho(t) = |z(t)|/\sigma$.

We trigger on the local maxima of each filter's time series when the \ac{SNR}
is above 5.5 and record those times and template parameters.  A list of
triggers is then passed to the next stage of the pipeline, which checks for
coincidence between detectors.

\subsubsection{Coincidence test}

We require events to be coincident in at least two detectors. For events to be
considered coincident, the time of coalescence and the masses of the
system\footnote{Strictly speaking, the chirp times $\tau_0$ and
$\tau_3$~\cite{hexabank}.} given by the triggers in each instrument must agree
to within a certain tolerance~\cite{Robinson:2008}.  Although we allowed for double coincident
combinations, we discarded H2L1 events that lacked an H1 trigger if H1 was
operating.  Since H1 was more sensitive than H2 it should have produced a
trigger for a real event.

As with the searches for low mass compact binary coalescences in \ac{S5}, we
used a coincidence test based on the template bank metric.  This test accounts
for correlations between the different parameters and attains a lower false
alarm rate for a given detection efficiency than simple parameter cuts.  As was
noted earlier, the metric used in this search was suboptimal. To take into
account this limitation, the coincidence requirements were looser than those of
previous \ac{S5} searches ~\cite{Collaboration:2009tt,Abbott:2009qj,
S5LowMassLV}. 

\subsubsection{Data quality vetoes}

Not all of the data taken during \ac{S5} was used for this analysis.  The
detectors frequently lost lock or were taken out of lock for commissioning
work.  Only times with stable lock stretches deemed as analyzable were marked
as \emph{science time}.  Segments of \emph{science time} containing more that
2048 s of data were analyzed in each of the three detectors H1, H2, L1.  

Occasionally data quality (DQ) during science time suffered from transient
excess noise.  Significant work was done to characterize these times prior to
examining the search candidates so as to not bias our detection and upper limit
statements~\cite{Slutsky:2010ff}.  Events at times suffering from poor data
quality are removed from the analysis.  The procedure of vetoing events reduces
the live time and also the false alarm rate of the search~\cite{Slutsky:2010ff}.
The following describes the basic procedure for vetoing candidates based on DQ.

The detectors are sensitive to a variety of noise transients (glitches) of
non-astrophysical origin, such as instrumental glitches and environmental
disturbances. The status of the detectors is monitored by a number of auxiliary
data channels that record the internal degrees of freedom of the
interferometers and the output from  environmental sensors. When the status of
a detector is suboptimal, the time is flagged.  Because the templates used in
this search have an impulse response lasting $\sim~10$ seconds, a short glitch
can produce triggers lasting several seconds after the glitch occurs. DQ flag
intervals often require search-specific time padding to improve the
effectiveness of the flag.  The length of this padding is determined by looking
at the distribution of triggers in the flagged interval.  The effectiveness of
a DQ flag is evaluated by the following metrics:
\emph{Efficiency:} the percentage of single detector triggers flagged.
Because these triggers are analyzed before coincidence, they
are dominated by transient noises local to the detector;
\emph {Dead-time:} the percentage of flagged time;
\emph{Used-percentage:} the fraction of flags that contain at least one
background trigger.
An effective flag has a high efficiency, a high used-percentage and a low
dead-time.  Flags found to be effective by these metrics are used as vetoes.

DQ flags are classified into four veto categories according to their metrics'
performance.  \emph{Category 1} contains times when the data was not analyzed
at all as described at the beginning of this section.  \emph{Category 2}
includes vetoes with a high efficiency-to-dead time ratio and a high used
percentage.  The origin of these glitches is well-understood and time intervals
are well-tuned.  \emph{Category 3} vetoes times with noise sources whose
coupling with the gravitational-wave channel is less understood, such as those
due to environmental noise. Category 3 vetoes are less correlated with
transients and are characterized by higher dead time and lower used percentage
than category 2 vetoes. Some flags, for example the overflows of digital
channels monitoring the alignment of the interferometer arm lengths and
mirrors, belong to both category 2 and 3 with different window lengths.
\emph{Category 4} contains vetoes with low efficiency and high dead time. These
flags usually identify minor environmental disturbances and problems recorded
in the electronic logbooks.

The DQ vetoes are used in the following way.  Category 2 vetoes are used
unconditionally in the search.  We examine events after Category 2 for
detection candidates.  However we apply Category 3 vetoes before creating the
list of candidates used to constrain the \ac{BBH} merger rate.  The category 3
veto list is chosen in advance in order to not bias our rate limit results.
Category 4 vetoes are used only to follow up interesting candidates, they do
not have any impact on the rate limits quoted in this paper.  All Category 2 or
greater vetoes are applied after the second coincidence stage before clustering
to produce the event list.  Vetoed time is accounted for to ensure that the
analyzed time calculations are correct.  Table \ref{t:times} gives the analyzed
time available after Category 3 vetoes are applied.
\begin{table}
\begin{center}
\begin{tabular}{ll}
\toprule
{\bf Detectors} \hspace{75 pt}  & {\bf Analyzed time (yr)} \hspace{45 pt} \\
\hline
 H1H2L1 & 0.6184\\
 H1L1 & 0.0968\\
 H2L1 & 0.0609\\
\bottomrule
\end{tabular}
\end{center}

\caption{The analyzed time surviving the pipeline after category 3 vetoes were
\label{t:times}
applied.  H1H2 times were not analyzed due to the inability to properly
estimate the background for co-located detectors.}
\end{table}

\subsection{Ranking and evaluation of candidate events}

\subsubsection{Signal consistency check}

Accounting for data quality as described in the previous section is not
sufficient to remove all triggers caused by environmental or instrumental
sources.  For that reason we employ a two stage pipeline that performs an
additional signal consistency check.  In the second filtering stage we
explicitly check the match of the signal to the template waveform by performing
a chi-squared test~\cite{Allen:2004}.  In this test, the template is divided
into $p$ frequency bins (for this search, we use 10 bins) such that each bin
contains the same expected contribution to the total SNR, if the signal matches
the template exactly. The SNR of the trigger in each bin is compared to the
expected SNR, and the differences are added in quadrature to obtain the value
of $\chi^2$.  We decompose the template waveforms into $p$ pieces of identical
power $\sigma^2 / p$
%
%
\begin{subequations}
\begin{align}
\tilde{h}(f) &= \sum_{i=1}^p \,\tilde{u}_i(f) , \\
\tilde{u}_i(f) &= \tilde{h}(f)
\, \Theta(f-f_{i,\text{low}}) \, \Theta(f_{i,\text{high}}-f)
\ .
\end{align}
\end{subequations}
Using (\ref{eq:filter}) we compute a filter time series for each of the
orthogonal pieces.
%
%
\begin{equation}
z_i(t) = 4 
            \int_0^{\infty} \frac{\tilde{u}_i(f)^* \tilde{s}(f)}
                      {S_n(f)} \, e^{2\pi i f t} \, d\!f .
\end{equation}
The $\chi^2$ statistic is then computed as
\begin{equation}
\chi^2(t) = \frac{1}{\sigma}\sum_{i=1}^p {\left| \frac{z(t)}{p} - z_i(t) \right|}^2 .
\end{equation}
Since $z(t)$is a complex number, corresponding to both phases
of the filter, the $\chi^2$ statistic has $2(p-1)$ degrees of freedom.

Previous searches in this mass range did not use \ac{IMR} waveforms.  Since the
models were not accurate they did not use a $\chi^2$ test~\cite{LIGOS3S4all}.
The $\chi^2$ statistic already provides significant separation from noise for a
large fraction of simulated signals in this search.  Future search efforts in
this mass range might employ new signal-based vetoes and multivariate
classifiers to achieve a better separation of signal from
background\footnote{In fact the analysis of the last six months of data in this
search used an additional discriminator called the $r^2$ veto, which checks the
consistency of the $\chi^2$ time series~\cite{Rodriguez:2007}}.

Once the $\chi^2$ statistic is evaluated we have almost all of the information
necessary to begin ranking events.  We describe in section \ref{sec:ranking}
how the $\chi^2$ statistic is folded together with the \ac{SNR} to produce a
ranking statistic known as effective \ac{SNR}.  First, however, in section
\ref{sec:back} we describe how we estimate our background, which is also
required for ranking the coincident events. 

\subsubsection{Background estimation\label{sec:back}}

We assume that instrumental noise triggers are not correlated between
detectors.  We estimate the background of this search by examining accidental
coincidences from time shifted data.  This section describes how we estimate
the background.  The next section describes how the background estimate is used
in ranking events.

In order to estimate the background of coincident events we repeat the
coincidence analysis with 100 time shifts between the two \ac{LIGO} sites in
multiples of five seconds.  We call the events found by this procedure time
slide events.  We expect that there will be no correlated noise between the
sites.  Therefore, the time-shifted analysis provides 100 background trials to
which we then compare the un-shifted data.  Unfortunately the assumption of
uncorrelated noise was not adequate for the collocated Hanford detectors, H1
and H2.  All events found in H1 and H2 but not L1 were discarded due to
correlated noise corrupting the background estimate. 

We find that the estimated background of the search is a function of time, the
parameters of the signals searched for and which detectors observed the event.
The total mass of the recovered signal is the best single parameter that
tracked the signal parameter dependence of the time slides.  We elaborate in
section \ref{sec:ranking} how this was used in the ranking of candidate events.

\subsubsection{Ranking events\label{sec:ranking}}

The ranking of candidate events is a multi-stage process.  The end ranking
statistic is a \ac{FAR} for each event that indicates how often
events like it (or louder than it) occur in time slides.  This section describes how we compute the
FAR and rank our events.

First, single-detector triggers are assigned an \emph{effective} \ac{SNR}
$\rho_{\text{eff}}$ which is a function of $\rho$ and $\chi^2$.  The
functional form is chosen to match the false alarm rate contours of the
single-detector background in the \ac{SNR} -- $\chi^2$ plane.  The effective
SNR is defined as
%
%
\begin{equation}
\rho_{\text{eff}} = \frac{\rho}{\left[(1+\rho^2/50)(\chi^2/\chi^2_{\text{dof}})\right]^{1/4}} ,
\end{equation}
where 50 is an empirically determined parameter and
$\chi^2_{\text{dof}}=2(p-1)$ is 18 for this search.  The single detector
effective SNRs $\rho_{\text{eff},i}$ are combined in quadrature to give a
coincident effective SNR
%
%
\begin{equation}
\rho_{\text{eff,c}} = \sqrt{\sum_i \rho_{\text{eff},i}^2} \;.
\label{eq:ceffsnr}
\end{equation} 

\begin{table}
  \centering
  \begin{tabular}{l|llll|l}
   \toprule
   {\bf Observation} \hspace{10 pt} & & {\bf Detectors} \hspace{15 pt} & {\bf Detectors} \hspace{15 pt} & & {\bf Mass} \\
   {\bf Epoch} & & {\bf Functioning, $F$} & {\bf Participating, $P$} & & {\bf Range, $M$} \\
   \hline
   $12 \sim$two-month & & H1L1 & H1L1 & & $[25,50)\, M_{\odot}$, \\
   epochs & & H2L1 & H2L1 & & $[50,85)\, M_{\odot}$, \\
      & & \multirow{2}{*}{H1H2L1} & H1H2L1 & & and \\
   \cline{4-4}
      & & & H1L1 & & $[85,100]\, M_{\odot}$ \\
   \bottomrule
 \end{tabular}
\caption{Breakdown of the analysis time and coincident trigger sets.  The
LIGO S5 run is divided into 12 epochs, each roughly two months in duration;
within each epoch, the time is divided according to which detectors were
operating and not vetoed.  Since there are three allowed combinations of
functioning detectors, there are $12\times 3=36$ different analyzed time
periods.  Different combinations of coincident events are allowed depending on
which detectors are functioning and participated in the coincident event.  There
are a total of four possible functioning/participating detector combinations
which contribute to the analysis.  Within each observation epoch and
functioning/participating detector combination, the events are divided into
three mass bins according to the average total mass of the templates involved
in the coincident event.  This means there are a total of $12\times 4\times
3=144$ different types of coincident events.  Each type of event has a separate
background distribution used to calculate its false-alarm probability.}
  \label{tab:breakdown}
\end{table}

We compute the FAR by comparing the un-shifted events to the time slide events.
Due to the non-Gaussian properties of the detector noise, the FAR depends on
the template.  It also depends on how many detectors were operating and
participated in the event.  We compute the FAR as a discrete function of four
parameters, the total mass $M$, the detectors that participate in the
coincidence $P$, the detectors that were functioning but not vetoed at the time
of the coincidence $F$, and the combined effective SNR rank of the event $R$.
We will denote a time-slide event that estimates our background as $B$. Each
parameter is an index for the event $B$.  The first and second indices, $F$ and $P$,
describe the instruments that were functioning during the event and the
detectors that participated in the event.  Only the following combinations were
considered: 1) triggers found in H1 and L1 when only the H1 and L1 detectors
were operating 2) triggers found in H1 and L1 when all three detectors H1, H2
and L1 were operating   3) triggers found in H2 and L1 when only H2 and L1 were
operating and 4) triggers found in all three detectors when all three detectors
were operating.  Note that as mentioned previously we were not able to estimate
a reliable background for triggers found only in H1 and H2.  Therefore those
events were discarded.  We also discarded events found in H2 and L1 when all
three detectors were on since the more sensitive H1 should observe a real
signal.  To summarize, the following shorthand notation for the 4 combinations
of participating $P$ and functioning $F$ detectors will be used:  $P,F \in
\{$H1L1,H1L1; H1L1,H1H2L1; H2L1,H2L1; H1H2L1,H1H2L1$\}$.  The third index $M$
denotes a range for the total mass estimated for the event and is in the set
$\{[25,50), [50,85), [85,100)\}M_{\odot}$.   The fourth index
$R$ is the rank of the event given by its effective SNR,
$\rho_{\text{eff,c}}$.  The $R$ index is determined by assigning the event
having a given $P$, $F$ and $M$ with the lowest combined effective SNR defined
in (\ref{eq:ceffsnr}) the value 0 and the the next lowest 1, etc., until all
events are ranked.  We calculate the false alarm rate $\FAR$ for a given
event as the number of all time slide events, $B$, with a rank ($R^+$) larger
than that event's rank divided by the time analyzed $T_F$ in the time shifted
analyses, which is a only a function of the instruments that were on and not
vetoed,
%
%
\begin{equation}
\FAR_{PFMR} = \sum_{R^+>R} B_{PFMR^+} \, T_F^{-1} .
\label{eq:FAR}
\end{equation}
This now allows us to map a zero-lag (unshifted) event to a $\FAR$ by
assigning it the same four parameters .

In addition to the indices describing how the false alarm rate was computed,
there is one remaining implicit parameter that refers to the time of the
events.  We separated the two calendar years of data into 12 two-month periods.
Each was treated separately for the calculation of \eqref{eq:FAR} in order to
crudely capture the variation of the noise properties over the course of
\ac{S5}.  It is worth making explicit the number of combinations over which
false alarm rates were computed.  Each of the 12 two-month periods had 4
possible combinations of detectors that were functioning and that produced
triggers as mentioned above.  Additionally each had 3 total mass bins.  The
result is $12 \times 4 \times 3 = 144$ separate calculations of \eqref{eq:FAR}.
This is described additionally in table \ref{tab:breakdown} and is relevant for
interpreting the significance of events in section \ref{sec:candidates}.

Next we assess the $\FAR$ of the events independently of the mass range $M$ and
the participating detectors $P$ in order to compute a global ranking that only
takes into account the detectors that were functioning and no other parameters.
To do this we use the inverse $\FAR^{-1}$ as an intermediate ranking
statistic to replace combined effective SNR as the rank in the index $R$.  We
denote these newly ranked time slide events as $B'$.  Then the combined $\FAR$
is 
%
%
\begin{equation}
\FAR_{FR} = \sum_{R^+>R}\sum_P\sum_M B'_{PFMR^+} \, T_F^{-1}
\ .
\label{eq:cFAR}
\end{equation}
The combined $\FAR$ is only a function of the detectors that were functioning
$F$ during the event and the inverse $\FAR$ rank computed at the previous step
$R$.  From the combined $\FAR$ we can also compute the False Alarm Probability
(FAP).  Assuming Poisson statistics, we define the FAP as the chance of getting
one or more events louder than the event in question purely from background.
This is defined as FAP$_{FR} = 1 - \exp{(-\FAR_{FR} T_F)}$.  $T_F$ is
nominally a particular detector combination live time for a two-month analysis
period, but can be replaced with the entire observation time in order to obtain
the FAP for an event given the result of all 12 two-month periods.


\section{Loudest coincident events}\label{sec:candidates}

As previously mentioned, we divided the $\sim2$ calendar years of data into 12
two-month blocks and this resulted in 144 separate computations of
\eqref{eq:FAR}.  Combining the $\FAR$ using \eqref{eq:cFAR} resulted in 36
separate periods consisting of distinct times when a given set of detectors were
functioning and providing data. These categories are independent since they
arise from distinct times.  Since a lower $\FAR$ implies a more significant event,
we use $\FAR^{-1}$ to rank the events.

Table \ref{t:loudest} gives the top 10 loudest events of the search ranked by
$\FAR^{-1}$.  Three of the 10 candidates were louder than any events in the 100
time-shifted coincident sets used to estimate the background.  The table
provides the bound on the $\FAR$ based on the total observed background time
during their two-month period when the same detectors were functioning.  We
note that it is not surprising to have events louder than the background given
the limitations of the background estimation. We used only 100 time shifts and
the number of trials examined for the computation of the $\FAR$ was 144.  We
therefore expected to observe $\sim 1.4$ events more significant than our
estimated background and we observed three.  In order to estimate the
significance of these three events we employed two additional techniques.
As the primary method, we
first interpolated and extrapolated the $\FAR$ from our 100 time-shift
background estimate.  To obtain an alternative estimate,
we extended our time-shift study to 1000 shifts
for the two-month periods in which those events occurred.  We decided before
un-blinding the analysis to use the extrapolated $\FAR$ values in the upper
limit computation when necessary.  We also examined many properties of these
events in a qualitative follow-up procedure.  The result of our analysis is
that all three events have FAPs of $> 10^{-2}$, assuming the full 0.8 yr
observation time, and all are consistent with rare instrumental noise
fluctuations; none are plausible candidate gravitational-wave detections. This
section provides some additional detail about these events.

\begin{table*}
\begin{center}
\begin{tabular}{llll|llll|llll|llll}
\toprule
{\bf Rank} \hspace{3 pt}  & {\bf FAR (yr$^{-1}$)} \hspace{3 pt}  & {\bf $\rho_{\mathrm{eff}}$} \hspace{3 pt}  & {\bf GPS Time} \hspace{20 pt}  & {\bf $\rho_{\mathrm{H1}}$} \hspace{5 pt}  & {\bf $\chi^2_{\mathrm{H1}}$} \hspace{5 pt}  & {\bf $m_{1\,\mathrm{H1}}$} \hspace{5 pt}  & {\bf $m_{2\,\mathrm{H1}}$} \hspace{5 pt}  & {\bf $\rho_{\mathrm{H2}}$} \hspace{5 pt}  & {\bf $\chi^2_{\mathrm{H2}}$} \hspace{5 pt}  & {\bf $m_{1\,\mathrm{H2}}$} \hspace{5 pt}  & {\bf $m_{2\,\mathrm{H2}}$} \hspace{5 pt}  & {\bf $\rho_{\mathrm{L1}}$} \hspace{5 pt}  & {\bf $\chi^2_{\mathrm{L1}}$} \hspace{5 pt}  & {\bf $m_{1\,\mathrm{L1}}$} \hspace{5 pt}  & {\bf $m_{2\,\mathrm{L1}}$} \hspace{5 pt} \\
\hline
 1 &  $<$ 0.20 &  12.8 &  848905672.3369 &  172.0 &  4057.9 &  94.0 &  6.0 &  24.4 &  167.4 &  49.7 &  17.3 &  8.3 &  46.0 &  95.2 & 4.8\\
 2 &  $<$ 0.25 &  11.6 &  825664840.1523 &  5.6 &  21.5 &  51.7 &  1.1 &  6.2 &  1.6 &  50.5 &  1.1 &  5.5 &  39.1 &  36.2 & 2.4\\
 3 &  $<$ 1.40 &  10.3 &  842749918.8057 &  - &  - &  - &  - &  5.5 &  7.8 &  67.1 &  2.5 &  12.2 &  20.4 &  83.2 & 16.8\\
 4 &  2.7 &  12.0 &  830222610.4062 &  5.5 &  34.2 &  98.0 &  2.0 &  * &  * &  * &  * &  28.8 &  43.5 &  91.7 & 8.3\\
 5 &  5.4 &  9.8 &  849056023.4121 &  5.7 &  11.3 &  29.9 &  1.3 &  * &  * &  * &  * &  5.6 &  2.5 &  23.6 & 1.8\\
 6 &  9.0 &  9.8 &  827865922.1265 &  9.2 &  67.6 &  37.3 &  1.2 &  - &  - &  - &  - &  6.8 &  4.2 &  31.0 & 1.5\\
 7 &  12 &  9.5 &  836048263.0366 &  6.2 &  15.0 &  52.9 &  1.4 &  6.4 &  14.6 &  46.7 &  1.6 &  5.9 &  21.3 &  53.0 & 1.3\\
 8 &  12 &  10.7 &  854487078.6543 &  6.1 &  29.6 &  96.7 &  3.3 &  * &  * &  * &  * &  18.1 &  29.8 &  97.0 & 3.0\\
 9 &  13 &  10.8 &  835998008.6890 &  23.2 &  52.3 &  94.8 &  5.2 &  * &  * &  * &  * &  5.8 &  21.2 &  78.1 & 1.2\\
 10 &  15 &  9.8 &  857817894.5767 &  8.8 &  29.7 &  90.7 &  1.4 &  9.9 &  40.8 &  94.8 &  5.2 &  5.9 &  28.1 &  90.4 & 1.4\\
\bottomrule
\end{tabular}
\end{center}

\caption{The loudest events of the search.  The coincident 
events are ranked by their combined false alarm rate $\FAR$. A `` - ''
represents that the detector was not functioning during the time of the event
in question.  A `` * '' represents that the detector was functioning but did
not produce a trigger above the single detector SNR threshold of 5.5.  Notice
that the top three events were found above their local background estimates.
For that reason only limits on their combined FARs are given here.  See the
text for details. 
\label{t:loudest}
}
\end{table*}

\subsection{H1H2L1 event at GPS time 848905672.3369\break 
(November 30, 2006 07:07:38.3369 GMT )}

The loudest event of this search at GPS time 848905672.3369 was found in all
three detectors and was more significant than any of the time-shifted events in
its background estimate. We put a bound on its $\FAR$ from the original 100
time-shift background estimate of 1 per 5 years.  We also estimated the $\FAR$
by interpolating and extrapolating the original 100 time-shift background
estimate using a fit to the trigger distribution.  The extrapolated $\FAR$ was
1/1.6\,yr.  Note that it was larger than the bound due to the fitting procedure
in the tail of the trigger distribution.  Also note that we decided in advance
to use the extrapolated FARs for the rate limit calculations in the next
section.  Therefore the FAR used for this event was 1/1.6\,yr.  We computed
1000 additional time shifts in this two-month period to better estimate its
false alarm probability.  From the additional time-shift background estimate we
computed that this event had a false alarm rate of 1/50\,yr.   

Given that we searched nearly 1 year of data, this event is consistent with
fluctuations.  The conservative probability of getting this event in background
(by choosing the lowest of the $\FAR$ estimates) is $\sim 0.02$. Our assessment
of this candidate is that it is a loud glitch in H1 with a moderate response in
H2 coincident with low amplitude noise in L1.  The ratio of distance estimates
associated with the signals in H1 and H2\footnote{Since H1 and H2 are colocated
and have the same antenna patterns, this ratio can be estimated independent of
geometrical effects.} is not consistent with a signal.  We measured a ratio of
$\sim 10$ and it should be $\sim 1$.  The H1 $\chi^2$ does not lie within the
expected signal distribution.  We therefore conclude that this is not a
gravitational-wave detection candidate.

\subsection{H1H2L1 event at GPS time 825664840.1523\break
(March 06, 2006 07:20:26.1523 GMT) \label{sec:event2}}

The second loudest event of this search at GPS time 825664840.1523 was more
significant than any of the 100 time slides performed during the two month
period in triple coincident H1,H2,L1 time.  The event was found in all three
detectors H1, H2, and L1 with SNR only slightly above threshold 5.60, 6.17 and
5.55 respectively.  The masses were consistent between the detectors.  In H1
and L1 this event had a $\chi^2$ that was consistent with both time slide
events and signals.  However, it had an unusually low $\chi^2$ value (0.1 per
degree of freedom) in H2.  A $\chi^2$ value of less than 0.1 per degree of
freedom is rare for both signals and noise.  No background events out of
$\sim300,000$ had such a low $\chi^2$ value nor did any of the $\sim10^6$
simulated signals.  The ranking of this event was artificially elevated by the
unusually low $\chi^2$ value.  If this event had a higher $\chi^2$ of 1 per
degree of freedom it would not stand above background. We conclude that this
event is not a gravitational-wave detection candidate.

The unusually low $\chi^2$ value put this event in a region of parameter space
where the FAR extrapolation is not valid.  This event happened
to occur in a segment of time that we reserved in advance as a test data set,
called a playground, that was not used in the rate limit calculation shown in
the next section.  See \ref{sec:upper limits} for more details. We place a
bound on its $\FAR$ of 1 per 4 years from the original 100 time-shift
background estimate.  We found that this candidate is not stable to small
changes in our analysis pipeline.  We were thus not able to measure its $\FAR$
independently using more time slides.   
 
\subsection{H2L1 event at GPS time 842749918.8057\break
(September 20, 2006 01:11:44.8057 GMT)}

The third loudest event of this search at GPS time 842749918.8057 was found in
H2 and L1. It was louder than any of the time slide events in its two-month
period.  We put a bound on its $\FAR$ of 1.4/\,yr from our original analysis.
An independent check using additional time slides yielded a $\FAR$ of
2.9/\,yr.  We also interpolated and extrapolated the original 100 time-shift
run to obtain a $\FAR$ of 1.9/\,yr. 

The $\FAR$ of 1.9/yr was used in the upper limit calculation described in the
next section. The L1 SNR and $\chi^2$ is consistent with the background in that
instrument.  The H2 trigger is just above the SNR threshold of 5.5.  This event
is not rare.  With a $\FAR$ of 1.9/yr we expected to observe an event similar to
this in our total observation time even though it was above background in its
local two-month H2L1 observation time.  We conclude that this event is not a
gravitational-wave detection candidate.


\section{Merger rate limits}\label{sec:upper limits}

Before examining events for detection candidates we agreed upon the procedure
described in this section for establishing an upper limit on the merger rate of
black hole binaries if no detections were found.  

In order to constrain the merger rate we had to assess the sensitivity of the
search.  To test the detection sensitivity of our search pipeline, we injected
$\sim10^6$ signals into the detector strain data and processed it with the same
pipeline used for the search.  Events associated with the injected signals
having FARs less than the loudest event of the search are considered to be
found by the pipeline.  We inject both \ac{EOB} and phenomenological waveforms
into the data. The injection parameters were as follows. For both waveform
families, the injected signals had distances between 1\,Mpc and 750\,Mpc
distributed uniformly in the logarithm of distance.  Both families had a
uniform distribution of sky location and orientation.  For both families the
total mass of the binary systems varied between $25 - 100\, \Msun$.  The
component mass distributions, however, did differ between the \ac{EOB} and
phenomenological waveforms.  The component mass distribution for \ac{EOB}
signals was generated by first producing a uniform distribution in the
component masses between $1 - 99\, \Msun$ and then clipping the distribution to
have no systems outside of the total mass range $25 - 100\, \Msun$.  The mass
distribution for the phenomenological waveforms was produced by first
generating a distribution that was uniform in mass ratio 
($m_1:m_2 \,;\, m_1 \geq m_2$) between $1:1$ and $10:1$ and then clipping the result to have no systems
outside of the total mass range $25 - 100\, \Msun$.

As previously stated we divided the $\sim 2$ years of data into 12 two-month
periods and examined each of the three functioning detector combinations
H1H2L1, H1L1, H2L1 separately for a total of 36 periods.  We reserved 10\% of
the detector time as an un-blinded \emph{playground}: we do not use playground
data in computing the upper limit on the merger rate.  The second loudest event
described in section \ref{sec:event2} happened to occur in the playground time. 
Using (\ref{eq:cFAR}) we ranked each candidate event in the 36 periods.  We
used the loudest event in the foreground after category 3 vetoes in each period
to establish a combined $\FAR$ threshold for determining what injections were
found.  For the events louder than background we used the extrapolated $\FAR$
as agreed on prior to un-blinding the analysis.   

\begin{figure}
\includegraphics[width=0.5\textwidth]{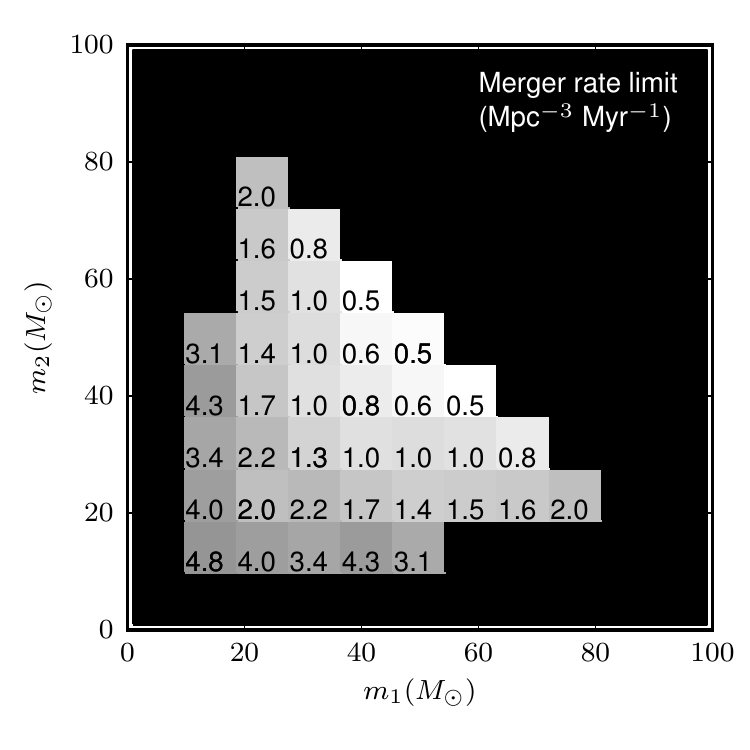}
\caption{\label{t:ul} The 90\% confidence upper limit on the merger rate as a
function of mass in units of\,{\Msun} (symmetric over $m_1$ and $m_2$).  This image
represents the rate limit in units of $\un{Mpc}^{-3}\un{Myr}^{-1}$.  These
limits can be converted to traditional units of $L_{10}^{-1}\un{Myr}^{-1}$ by
dividing by $0.0198\,L_{10}\un{Mpc}^{-3}$~\cite{LIGOS3S4Galaxies}.  Only bins
with mass ratios $< 4:1$ have upper limits computed due to uncertainty in the
waveform models for more asymmetric systems.}
\end{figure}

The efficiency $\bar{\epsilon}$ of recovering simulated signals in the
detection pipeline is a function of the loudest event $\FAR$, $\FAR^*$, the
radial distance to the source $r$ and the masses $m_1,m_2$.  Note that in
practice the mass dependence is captured by binning the mass plane into the
boxes illustrated in figure \ref{t:ul}  The bar denotes that the efficiency is
averaged over sky position and orientation.  We define the efficiency as
%
%
\begin{equation}
\bar{\epsilon}(\FAR^*, r, m_1, m_2) = 
       \frac{N_f(\FAR^*, r, m_1, m_2)}{N_t(\FAR^*, r, m_1, m_2)} ,
\end{equation}
where $N_f$ is the number of found injections, $N_t$ is the total number of
injections and $\FAR^*$ is the $\FAR$ of the loudest event in a given analysis
period.  We then compute the volume of the sky surveyed in each of the 36
independent observation periods (denoted by the index $i$) by
%
%
\begin{equation}
V_i(m_1,m_2,\FAR^*) 
= \int 4 \pi r^2 \bar{\epsilon_i}(\FAR^*, r, m_1, m_2) \,dr ,
\label{eq:volume}
\end{equation}
%
%
which has units of Mpc$^3$.  We estimate the variance
\begin{equation}
\label{eq:volumevar}
\sigma_i^2(m_1,m_2,\FAR^*)
= \langle V_i(m_1,m_2)^2 \rangle - {\langle V_i(m_1,m_2) \rangle}^2 
\end{equation}
by bootstrapping the input injection distribution to account for Monte-Carlo
errors as well as varying the injection distances according to the conservative
quadrature sum of the calibration uncertainty among the three detectors,
$20\%$~\cite{S5Calibration}.  An additional systematic error is associated with
uncertainty in the target waveforms.  These limits are presented with our best
understanding of the currently available waveforms.  If we take the fractional
difference in the SNRs of Phenomenological IMR and EOB waveforms, $\sim 10\%$
(see section \ref{sec:modelerrors}), as an indication of the uncertainty in the
range due to imperfectly known waveforms, we conclude that the rates as
reported in figure \ref{t:ul} have an additional systematic uncertainty of
$\sim 30\%$. This uncertainty is not included in the rate estimates nor are any
other systematic errors, for example the accuracy of the waveform phasing.
Some errors are discussed in~\cite{Buonanno:2007pf,Ajith:2007kx}.

In order to establish a merger rate $R(m_1,m_2)$ in units of mergers Mpc$^{-3}$
yr$^{-1}$  we adopt formula (24) in~\cite{Biswas:2007ni}.  It is important to
note that some simplification of these formulas occurs when choosing the $\FAR$
as the ranking statistic ~\cite{Keppel:thesis}.  Adapting the loudest
event formalism described in \cite{Biswas:2007ni} to our notation, if
we constructed a posterior on $R$ using only the results of a single
analysis period, the marginalized likelihood function would be
\begin{equation}
  \label{e:singlelikelihood}
  p(k_i,\Omega_i,\Lambda_i|R)
  \propto
  \left[
    \frac{1}{(1+R \Omega_i / k_i)^{k_i+1}} + 
    \frac{R \Omega_i \Lambda_i ( 1 + 1/k_i)}
    {(1+R \Omega_i / k_i)^{k_i+2}} 
  \right]
\end{equation}
where
\begin{align}
  \Omega_i &= V_i(m_1,m_2,\FAR^*) \, T_i,
  \\
  k_i &= {\left[\frac{V_i(m_1,m_2,\FAR^*)}{\sigma_i(m_1,m_2,\FAR^*)}\right]}^2 ,
  \\
  \Lambda_i &=
  \frac{\diff\ln{[V_i(m_1,m_2,\FAR^*)}]}{\diff\FAR^{*}}\frac{1}{T_i},
\end{align}
$T_i$ is the analyzed time for index $i$ (assumed to have no errors),
$V_i$ is taken from \eqref{eq:volume}, and the proportionality
constant in \eqref{e:singlelikelihood} can depend on $\Omega_i$, $k_i$ and
$\Lambda_i$, but not $R$.

In order to obtain the combined posterior probability distribution for
the rate, given the sensitivities and loudest events of the 36
different analysis periods, labeled by the index $i$, we multiply the
likelihood functions and assume an initial uniform prior on the
rate. This results in a posterior probability of the form
%
%
\begin{equation}
\label{eq:post}
  \begin{split}
    p(R|m_1,m_2) &\equiv p(R|\{k_i\},\{\Omega_i\},\{\Lambda_i\})
    \\
    &\propto
    p(\{k_i\},\{\Omega_i\},\{\Lambda_i\}|R)
    = \prod_i p(k_i,\Omega_i,\Lambda_i|R)    
  \end{split}
\end{equation}
We integrate the normalized form of \eqref{eq:post} to 90\% to establish the
90\% confidence upper limit on the merger rate (still a function of component
mass), $R_{90\%}$.  The result is given in figure \ref{t:ul}.  The upper limit
in the lowest mass bin considered in this search is an order of magnitude
higher than the most optimistic binary black hole merger rates predicted by
current population-synthesis studies (see, e.g.,
\cite{ratesdoc,Belczynski:2010,Bulik:2008}).   At the upper end of the analyzed
mass range, there are no reliable estimates for merger rates for intermediate
mass black holes, whose very existence remains to be confirmed; however, see
\cite{Mandel:2007rates,imbhlisa-2006,ratesdoc,Amaro:2006imbh} for some
intriguing possibilities.

\begin{figure}
\includegraphics[width=0.5\textwidth]{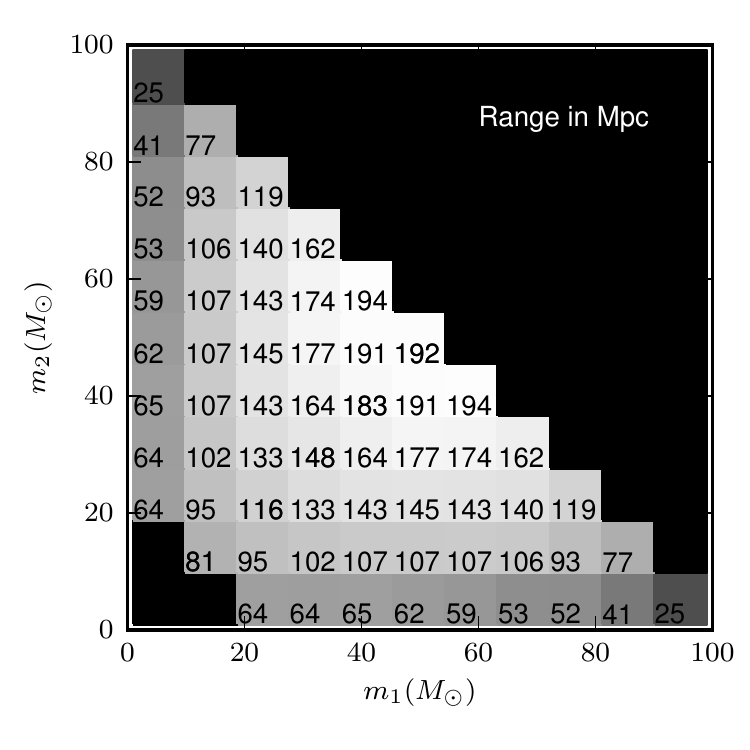}
\caption{Average range defined in \eqref{eq:range} for the search in Mpc as a
function of mass (symmetric over $m_1$ and $m_2$), assuming target waveforms that
match the EOB and Phenomenological models.}
\label{fig:range}
\end{figure}

As discussed above, due to the uncertainties in the waveform models for
asymmetric systems, we do not present upper limits for mass ratios $< 4:1$.
However, we do provide an average range for systems with smaller mass ratios
based on the EOB and Phenomenological waveform models, in figure
\ref{fig:range}.
The average range is defined as
%
%
\begin{subequations}
\label{eq:range}
\begin{align}
\langle \mathcal{R}(m_1,m_2,\FAR^*)\rangle & = \frac{1}{\sum_i T_i} \sum_i T_i \mathcal{R}_i(m_1,m_2,\FAR^*)
\ ,
\\
\mathcal{R}_i(m_1,m_2,\FAR^*) & = \left[\frac{3}{4\pi} V_i(m_1,m_2,\FAR^*) \right]^{1/3} ,
\end{align}
\end{subequations}
where $V_i(m_1,m_2,\FAR^*)$ is defined in \eqref{eq:volume}, $\mathcal{R}_i$ is
the radius of the sphere having volume of $V_i$ and the average range $\langle
\mathcal{R} \rangle$ is the time-weighted average of ranges computed from each
of the ranges found by examining the loudest event in each of the 36 periods.

\section{Conclusions}\label{sec:conclusions}
%
We presented the result of a search for \ac{BBH} coalescence during LIGO's
fifth science run spanning approximately two years of data taken from fall 2005
to fall 2007.  We targeted binaries with total mass $M=m_1+m_2$ in the range
$25\,{\Msun}\le M\le 100\,{\Msun}$ and component masses of $1\,{\Msun}\le
m_1,m_2 \le 99\,{\Msun}$ with negligible spin.  In order to effectually detect
such systems with LIGO it was necessary to use template waveforms that
encompass the inspiral, merger and ringdown phases of compact binary
coalescence.  We employed two waveform families in this search to filter and
assess the sensitivity.  Both had been tuned to numerical relativity
simulations.

%
We did not detect any plausible gravitational-wave candidates.  However we
estimated our search sensitivity and were able to constrain the merger rate of
the targeted sources in the nearby Universe.  We established to 90\% confidence
that the merger rate of black holes with component masses \ulrange is less than
\ulnum.  We note that this is still about an order of magnitude higher
than optimistic estimates for such systems~\cite{ratesdoc} (see also \cite{Bulik:2008,Belczynski:2010}).


There are a number of limitations in the current approach, which will be
addressed in future searches. The main limitation is that the template
waveforms neglect the effects of spin. Although the statistical distribution of
the spins of black holes in binaries is not well
known~\cite{MandelOShaughnessy:2010}, there are examples of black holes in
X-ray binaries which have been observed to have a large
spin~\cite{McClintock:2006xd}. For a binary with spinning components, the
expected observed gravitational-wave signal will differ from the non-spinning
case; the observed duration can be different and there may be modulation of the
gravitational-wave amplitude and phase.  Neglecting such effects in the search
templates will affect the detection efficiency for binaries with spinning
components.  However, at the time this search was conducted there were no
analytical inspiral-merger-ringdown waveforms for systems with generic spins.
This should be borne in mind when interpreting the results of the search.

Another limitation of the search is that, due to the shorter duration and
bandwidth of the signals in comparison to searches for lower mass systems, it
is harder to distinguish between genuine signals and background events, since
the signals themselves are more ``glitch-like''.  New approaches to the ranking
of candidate events are being developed to improve the sensitivity of searches
for these systems.

\acknowledgments

The authors gratefully acknowledge the support of the United States National
Science Foundation for the construction and operation of the LIGO Laboratory,
the Science and Technology Facilities Council of the United Kingdom, the
Max-Planck-Society, and the State of Niedersachsen/Germany for support of the
construction and operation of the GEO600 detector, and the Italian Istituto
Nazionale di Fisica Nucleare and the French Centre National de la Recherche
Scientifique for the construction and operation of the Virgo detector. The
authors also gratefully acknowledge the support of the research by these
agencies and by the Australian Research Council, the Council of Scientific and
Industrial Research of India, the Istituto Nazionale di Fisica Nucleare of
Italy, the Spanish Ministerio de Educaci\'on y Ciencia, the Conselleria
d'Economia Hisenda i Innovaci\'o of the Govern de les Illes Balears, the
Foundation for Fundamental Research on Matter supported by the Netherlands
Organisation for Scientific Research, the Polish Ministry of Science and Higher
Education, the FOCUS Programme of Foundation for Polish Science, the Royal
Society, the Scottish Funding Council, the Scottish Universities Physics
Alliance, The National Aeronautics and Space Administration, the Carnegie
Trust, the Leverhulme Trust, the David and Lucile Packard Foundation, the
Research Corporation, and the Alfred P. Sloan Foundation.

\appendix

\bibliography{s5-highMass}

\begin{thebibliography}{120}
\expandafter\ifx\csname natexlab\endcsname\relax\def\natexlab#1{#1}\fi
\expandafter\ifx\csname bibnamefont\endcsname\relax
  \def\bibnamefont#1{#1}\fi
\expandafter\ifx\csname bibfnamefont\endcsname\relax
  \def\bibfnamefont#1{#1}\fi
\expandafter\ifx\csname citenamefont\endcsname\relax
  \def\citenamefont#1{#1}\fi
\expandafter\ifx\csname url\endcsname\relax
  \def\url#1{\texttt{#1}}\fi
\expandafter\ifx\csname urlprefix\endcsname\relax\def\urlprefix{URL }\fi
\providecommand{\bibinfo}[2]{#2}
\providecommand{\eprint}[2][]{\url{#2}}

\bibitem[{\citenamefont{Abbott et~al.}(2009{\natexlab{a}})}]{Abbott:2007kv}
\bibinfo{author}{\bibfnamefont{B.}~\bibnamefont{Abbott}} \bibnamefont{et~al.}
  (\bibinfo{collaboration}{LIGO Scientific Collaboration}),
  \bibinfo{journal}{Rept.~Prog.~Phys.} \textbf{\bibinfo{volume}{72}},
  \bibinfo{pages}{076901} (\bibinfo{year}{2009}{\natexlab{a}}),
  \eprint{arXiv:0711.3041}.

\bibitem[{\citenamefont{{Acernese} et~al.}(2008)}]{Acernese:2008b}
\bibinfo{author}{\bibfnamefont{F.}~\bibnamefont{{Acernese}}}
  \bibnamefont{et~al.}, \bibinfo{journal}{Class. Quantum Grav.}
  \textbf{\bibinfo{volume}{25}}, \bibinfo{pages}{184001}
  (\bibinfo{year}{2008}).

\bibitem[{\citenamefont{Abbott
  et~al.}(2009{\natexlab{b}})}]{Collaboration:2009tt}
\bibinfo{author}{\bibfnamefont{B.}~\bibnamefont{Abbott}} \bibnamefont{et~al.}
  (\bibinfo{collaboration}{LIGO Scientific Collaboration}),
  \bibinfo{journal}{Phys.~Rev.~D} \textbf{\bibinfo{volume}{79}},
  \bibinfo{pages}{122001} (\bibinfo{year}{2009}{\natexlab{b}}),
  \eprint{arXiv:0901.0302}.

\bibitem[{\citenamefont{Abbott et~al.}(2009{\natexlab{c}})}]{Abbott:2009qj}
\bibinfo{author}{\bibfnamefont{B.}~\bibnamefont{Abbott}} \bibnamefont{et~al.}
  (\bibinfo{collaboration}{LIGO Scientific Collaboration}),
  \bibinfo{journal}{Phys.~Rev.~D} \textbf{\bibinfo{volume}{80}},
  \bibinfo{pages}{047101} (\bibinfo{year}{2009}{\natexlab{c}}).

\bibitem[{\citenamefont{Abadie et~al.}(2010{\natexlab{a}})}]{S5LowMassLV}
\bibinfo{author}{\bibfnamefont{J.}~\bibnamefont{Abadie}} \bibnamefont{et~al.}
  (\bibinfo{collaboration}{LIGO Scientific Collaboration and Virgo
  Collaboration}), \bibinfo{journal}{\prd} \textbf{\bibinfo{volume}{82}},
  \bibinfo{pages}{102001} (\bibinfo{year}{2010}{\natexlab{a}}),
  \eprint{1005.4655}.

\bibitem[{\citenamefont{Remillard and McClintock}(2006)}]{Remillard:2006fc}
\bibinfo{author}{\bibfnamefont{R.~A.} \bibnamefont{Remillard}}
  \bibnamefont{and} \bibinfo{author}{\bibfnamefont{J.~E.}
  \bibnamefont{McClintock}}, \bibinfo{journal}{Ann. Rev. Astron. Astrophys.}
  \textbf{\bibinfo{volume}{44}}, \bibinfo{pages}{49} (\bibinfo{year}{2006}),
  \eprint{astro-ph/0606352}.

\bibitem[{\citenamefont{{Orosz} et~al.}(2007)\citenamefont{{Orosz},
  {McClintock}, {Narayan}, {Bailyn}, {Hartman}, {Macri}, {Liu}, {Pietsch},
  {Remillard}, {Shporer} et~al.}}]{Orosz:2007}
\bibinfo{author}{\bibfnamefont{J.~A.} \bibnamefont{{Orosz}}},
  \bibinfo{author}{\bibfnamefont{J.~E.} \bibnamefont{{McClintock}}},
  \bibinfo{author}{\bibfnamefont{R.}~\bibnamefont{{Narayan}}},
  \bibinfo{author}{\bibfnamefont{C.~D.} \bibnamefont{{Bailyn}}},
  \bibinfo{author}{\bibfnamefont{J.~D.} \bibnamefont{{Hartman}}},
  \bibinfo{author}{\bibfnamefont{L.}~\bibnamefont{{Macri}}},
  \bibinfo{author}{\bibfnamefont{J.}~\bibnamefont{{Liu}}},
  \bibinfo{author}{\bibfnamefont{W.}~\bibnamefont{{Pietsch}}},
  \bibinfo{author}{\bibfnamefont{R.~A.} \bibnamefont{{Remillard}}},
  \bibinfo{author}{\bibfnamefont{A.}~\bibnamefont{{Shporer}}},
  \bibnamefont{et~al.}, \bibinfo{journal}{\nat} \textbf{\bibinfo{volume}{449}},
  \bibinfo{pages}{872} (\bibinfo{year}{2007}), \eprint{0710.3165}.

\bibitem[{\citenamefont{Ozel et~al.}(2010)\citenamefont{Ozel, Psaltis, Narayan,
  and McClintock}}]{Ozel:2010}
\bibinfo{author}{\bibfnamefont{F.}~\bibnamefont{Ozel}},
  \bibinfo{author}{\bibfnamefont{D.}~\bibnamefont{Psaltis}},
  \bibinfo{author}{\bibfnamefont{R.}~\bibnamefont{Narayan}}, \bibnamefont{and}
  \bibinfo{author}{\bibfnamefont{J.~E.} \bibnamefont{McClintock}},
  \bibinfo{journal}{Astrophys. J.} \textbf{\bibinfo{volume}{725}},
  \bibinfo{pages}{1918} (\bibinfo{year}{2010}), \eprint{1006.2834}.

\bibitem[{\citenamefont{{Farr} et~al.}(2010)\citenamefont{{Farr}, {Sravan},
  {Cantrell}, {Kreidberg}, {Bailyn}, {Mandel}, and {Kalogera}}}]{Farr:2010}
\bibinfo{author}{\bibfnamefont{W.~M.} \bibnamefont{{Farr}}},
  \bibinfo{author}{\bibfnamefont{N.}~\bibnamefont{{Sravan}}},
  \bibinfo{author}{\bibfnamefont{A.}~\bibnamefont{{Cantrell}}},
  \bibinfo{author}{\bibfnamefont{L.}~\bibnamefont{{Kreidberg}}},
  \bibinfo{author}{\bibfnamefont{C.~D.} \bibnamefont{{Bailyn}}},
  \bibinfo{author}{\bibfnamefont{I.}~\bibnamefont{{Mandel}}}, \bibnamefont{and}
  \bibinfo{author}{\bibfnamefont{V.}~\bibnamefont{{Kalogera}}},
  \bibinfo{journal}{ArXiv e-prints}  (\bibinfo{year}{2010}),
  \eprint{1011.1459}.

\bibitem[{\citenamefont{{Bulik} and
  {Belczy{\'n}ski}}(2003)}]{BulikBelczynski:2003}
\bibinfo{author}{\bibfnamefont{T.}~\bibnamefont{{Bulik}}} \bibnamefont{and}
  \bibinfo{author}{\bibfnamefont{K.}~\bibnamefont{{Belczy{\'n}ski}}},
  \bibinfo{journal}{\apjl} \textbf{\bibinfo{volume}{589}}, \bibinfo{pages}{L37}
  (\bibinfo{year}{2003}), \eprint{astro-ph/0301470}.

\bibitem[{\citenamefont{{O'Shaughnessy}
  et~al.}(2010)\citenamefont{{O'Shaughnessy}, {Kalogera}, and
  {Belczynski}}}]{OShaughnessy:2009}
\bibinfo{author}{\bibfnamefont{R.}~\bibnamefont{{O'Shaughnessy}}},
  \bibinfo{author}{\bibfnamefont{V.}~\bibnamefont{{Kalogera}}},
  \bibnamefont{and}
  \bibinfo{author}{\bibfnamefont{K.}~\bibnamefont{{Belczynski}}},
  \bibinfo{journal}{\apj} \textbf{\bibinfo{volume}{716}}, \bibinfo{pages}{615}
  (\bibinfo{year}{2010}), \eprint{0908.3635}.

\bibitem[{\citenamefont{{Crowther} et~al.}(2010)\citenamefont{{Crowther},
  {Barnard}, {Carpano}, {Clark}, {Dhillon}, and {Pollock}}}]{Crowther:2010}
\bibinfo{author}{\bibfnamefont{P.~A.} \bibnamefont{{Crowther}}},
  \bibinfo{author}{\bibfnamefont{R.}~\bibnamefont{{Barnard}}},
  \bibinfo{author}{\bibfnamefont{S.}~\bibnamefont{{Carpano}}},
  \bibinfo{author}{\bibfnamefont{J.~S.} \bibnamefont{{Clark}}},
  \bibinfo{author}{\bibfnamefont{V.~S.} \bibnamefont{{Dhillon}}},
  \bibnamefont{and} \bibinfo{author}{\bibfnamefont{A.~M.~T.}
  \bibnamefont{{Pollock}}}, \bibinfo{journal}{\mnras} pp. \bibinfo{pages}{L11+}
  (\bibinfo{year}{2010}), \eprint{1001.4616}.

\bibitem[{\citenamefont{{Bulik} et~al.}(2008)\citenamefont{{Bulik},
  {Belczynski}, and {Prestwich}}}]{Bulik:2008}
\bibinfo{author}{\bibfnamefont{T.}~\bibnamefont{{Bulik}}},
  \bibinfo{author}{\bibfnamefont{K.}~\bibnamefont{{Belczynski}}},
  \bibnamefont{and}
  \bibinfo{author}{\bibfnamefont{A.}~\bibnamefont{{Prestwich}}},
  \bibinfo{journal}{ArXiv e-prints}  (\bibinfo{year}{2008}),
  \eprint{0803.3516}.

\bibitem[{\citenamefont{{O'Leary} et~al.}(2007)\citenamefont{{O'Leary},
  {O'Shaughnessy}, and {Rasio}}}]{OLeary:2007}
\bibinfo{author}{\bibfnamefont{R.~M.} \bibnamefont{{O'Leary}}},
  \bibinfo{author}{\bibfnamefont{R.}~\bibnamefont{{O'Shaughnessy}}},
  \bibnamefont{and} \bibinfo{author}{\bibfnamefont{F.~A.}
  \bibnamefont{{Rasio}}}, \bibinfo{journal}{Phys.~Rev.~D}
  \textbf{\bibinfo{volume}{76}}, \bibinfo{pages}{061504}
  (\bibinfo{year}{2007}), \eprint{astro-ph/0701887}.

\bibitem[{\citenamefont{{Sadowski} et~al.}(2008)\citenamefont{{Sadowski},
  {Belczynski}, {Bulik}, {Ivanova}, {Rasio}, and {O'Shaughnessy}}}]{Sadowski}
\bibinfo{author}{\bibfnamefont{A.}~\bibnamefont{{Sadowski}}},
  \bibinfo{author}{\bibfnamefont{K.}~\bibnamefont{{Belczynski}}},
  \bibinfo{author}{\bibfnamefont{T.}~\bibnamefont{{Bulik}}},
  \bibinfo{author}{\bibfnamefont{N.}~\bibnamefont{{Ivanova}}},
  \bibinfo{author}{\bibfnamefont{F.~A.} \bibnamefont{{Rasio}}},
  \bibnamefont{and}
  \bibinfo{author}{\bibfnamefont{R.}~\bibnamefont{{O'Shaughnessy}}},
  \bibinfo{journal}{\apj} \textbf{\bibinfo{volume}{676}}, \bibinfo{pages}{1162}
  (\bibinfo{year}{2008}), \eprint{0710.0878}.

\bibitem[{\citenamefont{{Miller} and {Lauburg}}(2009)}]{MillerLauburg:2008}
\bibinfo{author}{\bibfnamefont{M.~C.} \bibnamefont{{Miller}}} \bibnamefont{and}
  \bibinfo{author}{\bibfnamefont{V.~M.} \bibnamefont{{Lauburg}}},
  \bibinfo{journal}{\apj} \textbf{\bibinfo{volume}{692}}, \bibinfo{pages}{917}
  (\bibinfo{year}{2009}), \eprint{0804.2783}.

\bibitem[{\citenamefont{{O'Leary} et~al.}(2009)\citenamefont{{O'Leary},
  {Kocsis}, and {Loeb}}}]{OLeary:2008}
\bibinfo{author}{\bibfnamefont{R.~M.} \bibnamefont{{O'Leary}}},
  \bibinfo{author}{\bibfnamefont{B.}~\bibnamefont{{Kocsis}}}, \bibnamefont{and}
  \bibinfo{author}{\bibfnamefont{A.}~\bibnamefont{{Loeb}}},
  \bibinfo{journal}{\mnras} \textbf{\bibinfo{volume}{395}},
  \bibinfo{pages}{2127} (\bibinfo{year}{2009}), \eprint{0807.2638}.

\bibitem[{\citenamefont{Belczynski et~al.}(2010)}]{Belczynski:2009}
\bibinfo{author}{\bibfnamefont{K.}~\bibnamefont{Belczynski}}
  \bibnamefont{et~al.}, \bibinfo{journal}{Astrophys. J.}
  \textbf{\bibinfo{volume}{714}}, \bibinfo{pages}{1217} (\bibinfo{year}{2010}),
  \eprint{0904.2784}.

\bibitem[{\citenamefont{{Belczynski} et~al.}(2010)\citenamefont{{Belczynski},
  {Dominik}, {Bulik}, {O'Shaughnessy}, {Fryer}, and {Holz}}}]{Belczynski:2010}
\bibinfo{author}{\bibfnamefont{K.}~\bibnamefont{{Belczynski}}},
  \bibinfo{author}{\bibfnamefont{M.}~\bibnamefont{{Dominik}}},
  \bibinfo{author}{\bibfnamefont{T.}~\bibnamefont{{Bulik}}},
  \bibinfo{author}{\bibfnamefont{R.}~\bibnamefont{{O'Shaughnessy}}},
  \bibinfo{author}{\bibfnamefont{C.}~\bibnamefont{{Fryer}}}, \bibnamefont{and}
  \bibinfo{author}{\bibfnamefont{D.~E.} \bibnamefont{{Holz}}},
  \bibinfo{journal}{\apjl} \textbf{\bibinfo{volume}{715}},
  \bibinfo{pages}{L138} (\bibinfo{year}{2010}), \eprint{1004.0386}.

\bibitem[{\citenamefont{{Miller} and {Colbert}}(2004)}]{2004IJMPD..13....1M}
\bibinfo{author}{\bibfnamefont{M.~C.} \bibnamefont{{Miller}}} \bibnamefont{and}
  \bibinfo{author}{\bibfnamefont{E.~J.~M.} \bibnamefont{{Colbert}}},
  \bibinfo{journal}{International Journal of Modern Physics D}
  \textbf{\bibinfo{volume}{13}}, \bibinfo{pages}{1} (\bibinfo{year}{2004}),
  \eprint{astro-ph/0308402}.

\bibitem[{\citenamefont{Miller}(2009)}]{miller-2008}
\bibinfo{author}{\bibfnamefont{M.~C.} \bibnamefont{Miller}},
  \bibinfo{journal}{Class. Quant. Grav.} \textbf{\bibinfo{volume}{26}},
  \bibinfo{pages}{094031} (\bibinfo{year}{2009}), \eprint{0812.3028}.

\bibitem[{\citenamefont{{Farrell} et~al.}(2009)\citenamefont{{Farrell}, {Webb},
  {Barret}, {Godet}, and {Rodrigues}}}]{2009Natur.460...73F}
\bibinfo{author}{\bibfnamefont{S.~A.} \bibnamefont{{Farrell}}},
  \bibinfo{author}{\bibfnamefont{N.~A.} \bibnamefont{{Webb}}},
  \bibinfo{author}{\bibfnamefont{D.}~\bibnamefont{{Barret}}},
  \bibinfo{author}{\bibfnamefont{O.}~\bibnamefont{{Godet}}}, \bibnamefont{and}
  \bibinfo{author}{\bibfnamefont{J.~M.} \bibnamefont{{Rodrigues}}},
  \bibinfo{journal}{\nat} \textbf{\bibinfo{volume}{460}}, \bibinfo{pages}{73}
  (\bibinfo{year}{2009}).

\bibitem[{\citenamefont{{Brown} et~al.}(2007)\citenamefont{{Brown}, {Brink},
  {Fang}, {Gair}, {Li}, {Lovelace}, {Mandel}, and {Thorne}}}]{Brown:2007}
\bibinfo{author}{\bibfnamefont{D.~A.} \bibnamefont{{Brown}}},
  \bibinfo{author}{\bibfnamefont{J.}~\bibnamefont{{Brink}}},
  \bibinfo{author}{\bibfnamefont{H.}~\bibnamefont{{Fang}}},
  \bibinfo{author}{\bibfnamefont{J.~R.} \bibnamefont{{Gair}}},
  \bibinfo{author}{\bibfnamefont{C.}~\bibnamefont{{Li}}},
  \bibinfo{author}{\bibfnamefont{G.}~\bibnamefont{{Lovelace}}},
  \bibinfo{author}{\bibfnamefont{I.}~\bibnamefont{{Mandel}}}, \bibnamefont{and}
  \bibinfo{author}{\bibfnamefont{K.~S.} \bibnamefont{{Thorne}}},
  \bibinfo{journal}{Phys. Rev. Lett.} \textbf{\bibinfo{volume}{99}},
  \bibinfo{pages}{201102} (\bibinfo{year}{2007}), \eprint{arXiv:gr-qc/0612060}.

\bibitem[{\citenamefont{{Mandel} et~al.}(2008)\citenamefont{{Mandel}, {Brown},
  {Gair}, and {Miller}}}]{Mandel:2007rates}
\bibinfo{author}{\bibfnamefont{I.}~\bibnamefont{{Mandel}}},
  \bibinfo{author}{\bibfnamefont{D.~A.} \bibnamefont{{Brown}}},
  \bibinfo{author}{\bibfnamefont{J.~R.} \bibnamefont{{Gair}}},
  \bibnamefont{and} \bibinfo{author}{\bibfnamefont{M.~C.}
  \bibnamefont{{Miller}}}, \bibinfo{journal}{\apj}
  \textbf{\bibinfo{volume}{681}}, \bibinfo{pages}{1431} (\bibinfo{year}{2008}),
  \eprint{0705.0285}.

\bibitem[{\citenamefont{{Fregeau} et~al.}(2006)\citenamefont{{Fregeau},
  {Larson}, {Miller}, {O'Shaughnessy}, and {Rasio}}}]{imbhlisa-2006}
\bibinfo{author}{\bibfnamefont{J.~M.} \bibnamefont{{Fregeau}}},
  \bibinfo{author}{\bibfnamefont{S.~L.} \bibnamefont{{Larson}}},
  \bibinfo{author}{\bibfnamefont{M.~C.} \bibnamefont{{Miller}}},
  \bibinfo{author}{\bibfnamefont{R.}~\bibnamefont{{O'Shaughnessy}}},
  \bibnamefont{and} \bibinfo{author}{\bibfnamefont{F.~A.}
  \bibnamefont{{Rasio}}}, \bibinfo{journal}{\apjl}
  \textbf{\bibinfo{volume}{646}}, \bibinfo{pages}{L135} (\bibinfo{year}{2006}),
  \eprint{astro-ph/0605732}.

\bibitem[{\citenamefont{{Amaro-Seoane} and {Freitag}}(2006)}]{Amaro:2006imbh}
\bibinfo{author}{\bibfnamefont{P.}~\bibnamefont{{Amaro-Seoane}}}
  \bibnamefont{and}
  \bibinfo{author}{\bibfnamefont{M.}~\bibnamefont{{Freitag}}},
  \bibinfo{journal}{\apjl} \textbf{\bibinfo{volume}{653}}, \bibinfo{pages}{L53}
  (\bibinfo{year}{2006}), \eprint{astro-ph/0610478}.

\bibitem[{\citenamefont{{Belczynski} et~al.}(2004)\citenamefont{{Belczynski},
  {Bulik}, and {Rudak}}}]{Belczynski:2004popIII}
\bibinfo{author}{\bibfnamefont{K.}~\bibnamefont{{Belczynski}}},
  \bibinfo{author}{\bibfnamefont{T.}~\bibnamefont{{Bulik}}}, \bibnamefont{and}
  \bibinfo{author}{\bibfnamefont{B.}~\bibnamefont{{Rudak}}},
  \bibinfo{journal}{\apjl} \textbf{\bibinfo{volume}{608}}, \bibinfo{pages}{L45}
  (\bibinfo{year}{2004}), \eprint{arXiv:astro-ph/0403361}.

\bibitem[{\citenamefont{Abadie et~al.}(2010{\natexlab{b}})}]{ratesdoc}
\bibinfo{author}{\bibfnamefont{J.}~\bibnamefont{Abadie}} \bibnamefont{et~al.}
  (\bibinfo{collaboration}{LIGO Scientific Collaboration and Virgo
  Collaboration}), \bibinfo{journal}{Class. Quantum Grav.}
  \textbf{\bibinfo{volume}{27}}, \bibinfo{pages}{173001}
  (\bibinfo{year}{2010}{\natexlab{b}}).

\bibitem[{\citenamefont{{Hahn} and {Lindquist}}(1964)}]{Hahn64}
\bibinfo{author}{\bibfnamefont{S.~G.} \bibnamefont{{Hahn}}} \bibnamefont{and}
  \bibinfo{author}{\bibfnamefont{R.~W.} \bibnamefont{{Lindquist}}},
  \bibinfo{journal}{Annals of Physics} \textbf{\bibinfo{volume}{29}},
  \bibinfo{pages}{304} (\bibinfo{year}{1964}).

\bibitem[{\citenamefont{Pretorius}(2005)}]{Pretorius:2005gq}
\bibinfo{author}{\bibfnamefont{F.}~\bibnamefont{Pretorius}},
  \bibinfo{journal}{Phys. Rev. Lett.} \textbf{\bibinfo{volume}{95}},
  \bibinfo{pages}{121101} (\bibinfo{year}{2005}), \eprint{gr-qc/0507014}.

\bibitem[{\citenamefont{Campanelli
  et~al.}(2006{\natexlab{a}})\citenamefont{Campanelli, Lousto, Marronetti, and
  Zlochower}}]{Campanelli:2005dd}
\bibinfo{author}{\bibfnamefont{M.}~\bibnamefont{Campanelli}},
  \bibinfo{author}{\bibfnamefont{C.~O.} \bibnamefont{Lousto}},
  \bibinfo{author}{\bibfnamefont{P.}~\bibnamefont{Marronetti}},
  \bibnamefont{and}
  \bibinfo{author}{\bibfnamefont{Y.}~\bibnamefont{Zlochower}},
  \bibinfo{journal}{Phys. Rev. Lett.} \textbf{\bibinfo{volume}{96}},
  \bibinfo{pages}{111101} (\bibinfo{year}{2006}{\natexlab{a}}),
  \eprint{gr-qc/0511048}.

\bibitem[{\citenamefont{Baker et~al.}(2006{\natexlab{a}})\citenamefont{Baker,
  Centrella, Choi, Koppitz, and van Meter}}]{Baker:2005vv}
\bibinfo{author}{\bibfnamefont{J.~G.} \bibnamefont{Baker}},
  \bibinfo{author}{\bibfnamefont{J.}~\bibnamefont{Centrella}},
  \bibinfo{author}{\bibfnamefont{D.-I.} \bibnamefont{Choi}},
  \bibinfo{author}{\bibfnamefont{M.}~\bibnamefont{Koppitz}}, \bibnamefont{and}
  \bibinfo{author}{\bibfnamefont{J.}~\bibnamefont{van Meter}},
  \bibinfo{journal}{Phys. Rev. Lett.} \textbf{\bibinfo{volume}{96}},
  \bibinfo{pages}{111102} (\bibinfo{year}{2006}{\natexlab{a}}),
  \eprint{gr-qc/0511103}.

\bibitem[{\citenamefont{Pretorius}(2009)}]{Pretorius:2007nq}
\bibinfo{author}{\bibfnamefont{F.}~\bibnamefont{Pretorius}}, in
  \emph{\bibinfo{booktitle}{Physics of Relativistic Objects in Compact
  Binaries: from Birth to Coalescence}}, edited by
  \bibinfo{editor}{\bibfnamefont{M.}~\bibnamefont{Colpi}},
  \bibinfo{editor}{\bibfnamefont{P.}~\bibnamefont{Casella}},
  \bibinfo{editor}{\bibfnamefont{V.}~\bibnamefont{Gorini}},
  \bibinfo{editor}{\bibfnamefont{U.}~\bibnamefont{Moschella}},
  \bibnamefont{and} \bibinfo{editor}{\bibfnamefont{A.}~\bibnamefont{Possenti}}
  (\bibinfo{publisher}{Springer}, \bibinfo{address}{Heidelberg, Germany},
  \bibinfo{year}{2009}), \eprint{arXiv:0710.1338}.

\bibitem[{\citenamefont{Husa}(2007)}]{Husa:2007zz}
\bibinfo{author}{\bibfnamefont{S.}~\bibnamefont{Husa}}, \bibinfo{journal}{Eur.
  Phys. J. ST} \textbf{\bibinfo{volume}{152}}, \bibinfo{pages}{183}
  (\bibinfo{year}{2007}), \eprint{0812.4395}.

\bibitem[{\citenamefont{Hannam}(2009)}]{Hannam:2009rd}
\bibinfo{author}{\bibfnamefont{M.}~\bibnamefont{Hannam}},
  \bibinfo{journal}{Class. Quantum Grav.} \textbf{\bibinfo{volume}{26}},
  \bibinfo{pages}{114001} (\bibinfo{year}{2009}), \eprint{arXiv:0901.2931}.

\bibitem[{\citenamefont{Hinder}(2010)}]{Hinder:2010vn}
\bibinfo{author}{\bibfnamefont{I.}~\bibnamefont{Hinder}},
  \bibinfo{journal}{Class. Quant. Grav.} \textbf{\bibinfo{volume}{27}},
  \bibinfo{pages}{114004} (\bibinfo{year}{2010}), \eprint{1001.5161}.

\bibitem[{\citenamefont{Aylott et~al.}(2009)}]{Aylott:2009ya}
\bibinfo{author}{\bibfnamefont{B.}~\bibnamefont{Aylott}} \bibnamefont{et~al.},
  \bibinfo{journal}{Class. Quantum Grav.} \textbf{\bibinfo{volume}{26}},
  \bibinfo{pages}{165008} (\bibinfo{year}{2009}), \eprint{arXiv:0901.4399}.

\bibitem[{\citenamefont{Herrmann
  et~al.}(2007{\natexlab{a}})\citenamefont{Herrmann, Hinder, Shoemaker, and
  Laguna}}]{Herrmann:2006ks}
\bibinfo{author}{\bibfnamefont{F.}~\bibnamefont{Herrmann}},
  \bibinfo{author}{\bibfnamefont{I.}~\bibnamefont{Hinder}},
  \bibinfo{author}{\bibfnamefont{D.}~\bibnamefont{Shoemaker}},
  \bibnamefont{and} \bibinfo{author}{\bibfnamefont{P.}~\bibnamefont{Laguna}},
  \bibinfo{journal}{Class. Quantum Grav.}
  (\bibinfo{year}{2007}{\natexlab{a}}), \eprint{gr-qc/0601026}.

\bibitem[{\citenamefont{Baker et~al.}(2006{\natexlab{b}})}]{Baker:2006vn}
\bibinfo{author}{\bibfnamefont{J.~G.} \bibnamefont{Baker}}
  \bibnamefont{et~al.}, \bibinfo{journal}{Astrophys. J.}
  \textbf{\bibinfo{volume}{653}}, \bibinfo{pages}{L93}
  (\bibinfo{year}{2006}{\natexlab{b}}), \eprint{astro-ph/0603204}.

\bibitem[{\citenamefont{Gonzalez
  et~al.}(2007{\natexlab{a}})\citenamefont{Gonzalez, Sperhake, Br\"{u}gmann,
  Hannam, and Husa}}]{Gonzalez:2006md}
\bibinfo{author}{\bibfnamefont{J.~A.} \bibnamefont{Gonzalez}},
  \bibinfo{author}{\bibfnamefont{U.}~\bibnamefont{Sperhake}},
  \bibinfo{author}{\bibfnamefont{B.}~\bibnamefont{Br\"{u}gmann}},
  \bibinfo{author}{\bibfnamefont{M.}~\bibnamefont{Hannam}}, \bibnamefont{and}
  \bibinfo{author}{\bibfnamefont{S.}~\bibnamefont{Husa}},
  \bibinfo{journal}{Phys. Rev. Lett.} \textbf{\bibinfo{volume}{98}},
  \bibinfo{pages}{091101} (\bibinfo{year}{2007}{\natexlab{a}}),
  \eprint{gr-qc/0610154}.

\bibitem[{\citenamefont{Herrmann
  et~al.}(2007{\natexlab{b}})\citenamefont{Herrmann, Hinder, Shoemaker, Laguna,
  and Matzner}}]{Herrmann:2007ac}
\bibinfo{author}{\bibfnamefont{F.}~\bibnamefont{Herrmann}},
  \bibinfo{author}{\bibfnamefont{I.}~\bibnamefont{Hinder}},
  \bibinfo{author}{\bibfnamefont{D.}~\bibnamefont{Shoemaker}},
  \bibinfo{author}{\bibfnamefont{P.}~\bibnamefont{Laguna}}, \bibnamefont{and}
  \bibinfo{author}{\bibfnamefont{R.~A.} \bibnamefont{Matzner}},
  \bibinfo{journal}{Astrophys. J.} \textbf{\bibinfo{volume}{661}},
  \bibinfo{pages}{430} (\bibinfo{year}{2007}{\natexlab{b}}),
  \eprint{gr-qc/0701143}.

\bibitem[{\citenamefont{Koppitz et~al.}(2007)}]{Koppitz:2007ev}
\bibinfo{author}{\bibfnamefont{M.}~\bibnamefont{Koppitz}} \bibnamefont{et~al.},
  \bibinfo{journal}{Phys. Rev. Lett.} \textbf{\bibinfo{volume}{99}},
  \bibinfo{pages}{041102} (\bibinfo{year}{2007}), \eprint{gr-qc/0701163}.

\bibitem[{\citenamefont{Campanelli
  et~al.}(2007{\natexlab{a}})\citenamefont{Campanelli, Lousto, Zlochower, and
  Merritt}}]{Campanelli:2007ew}
\bibinfo{author}{\bibfnamefont{M.}~\bibnamefont{Campanelli}},
  \bibinfo{author}{\bibfnamefont{C.~O.} \bibnamefont{Lousto}},
  \bibinfo{author}{\bibfnamefont{Y.}~\bibnamefont{Zlochower}},
  \bibnamefont{and} \bibinfo{author}{\bibfnamefont{D.}~\bibnamefont{Merritt}},
  \bibinfo{journal}{Astrophys. J.} \textbf{\bibinfo{volume}{659}},
  \bibinfo{pages}{L5} (\bibinfo{year}{2007}{\natexlab{a}}),
  \eprint{gr-qc/0701164}.

\bibitem[{\citenamefont{Gonzalez
  et~al.}(2007{\natexlab{b}})\citenamefont{Gonzalez, Hannam, Sperhake,
  Br\"{u}gmann, and Husa}}]{Gonzalez:2007hi}
\bibinfo{author}{\bibfnamefont{J.~A.} \bibnamefont{Gonzalez}},
  \bibinfo{author}{\bibfnamefont{M.~D.} \bibnamefont{Hannam}},
  \bibinfo{author}{\bibfnamefont{U.}~\bibnamefont{Sperhake}},
  \bibinfo{author}{\bibfnamefont{B.}~\bibnamefont{Br\"{u}gmann}},
  \bibnamefont{and} \bibinfo{author}{\bibfnamefont{S.}~\bibnamefont{Husa}},
  \bibinfo{journal}{Phys. Rev. Lett.} \textbf{\bibinfo{volume}{98}},
  \bibinfo{pages}{231101} (\bibinfo{year}{2007}{\natexlab{b}}),
  \eprint{gr-qc/0702052}.

\bibitem[{\citenamefont{Tichy and Marronetti}(2007)}]{Tichy:2007hk}
\bibinfo{author}{\bibfnamefont{W.}~\bibnamefont{Tichy}} \bibnamefont{and}
  \bibinfo{author}{\bibfnamefont{P.}~\bibnamefont{Marronetti}},
  \bibinfo{journal}{Phys. Rev. D} \textbf{\bibinfo{volume}{76}},
  \bibinfo{pages}{061502} (\bibinfo{year}{2007}), \eprint{gr-qc/0703075}.

\bibitem[{\citenamefont{Campanelli
  et~al.}(2007{\natexlab{b}})\citenamefont{Campanelli, Lousto, Zlochower, and
  Merritt}}]{Campanelli:2007cga}
\bibinfo{author}{\bibfnamefont{M.}~\bibnamefont{Campanelli}},
  \bibinfo{author}{\bibfnamefont{C.~O.} \bibnamefont{Lousto}},
  \bibinfo{author}{\bibfnamefont{Y.}~\bibnamefont{Zlochower}},
  \bibnamefont{and} \bibinfo{author}{\bibfnamefont{D.}~\bibnamefont{Merritt}},
  \bibinfo{journal}{Phys. Rev. Lett.} \textbf{\bibinfo{volume}{98}},
  \bibinfo{pages}{231102} (\bibinfo{year}{2007}{\natexlab{b}}),
  \eprint{gr-qc/0702133}.

\bibitem[{\citenamefont{Baker et~al.}(2007{\natexlab{a}})}]{Baker:2007gi}
\bibinfo{author}{\bibfnamefont{J.~G.} \bibnamefont{Baker}}
  \bibnamefont{et~al.}, \bibinfo{journal}{Astrophys. J.}
  \textbf{\bibinfo{volume}{668}}, \bibinfo{pages}{1140}
  (\bibinfo{year}{2007}{\natexlab{a}}), \bibinfo{note}{astro-ph/0702390}.

\bibitem[{\citenamefont{Herrmann
  et~al.}(2007{\natexlab{c}})\citenamefont{Herrmann, Hinder, Shoemaker, Laguna,
  and Matzner}}]{Herrmann:2007ex}
\bibinfo{author}{\bibfnamefont{F.}~\bibnamefont{Herrmann}},
  \bibinfo{author}{\bibfnamefont{I.}~\bibnamefont{Hinder}},
  \bibinfo{author}{\bibfnamefont{D.~M.} \bibnamefont{Shoemaker}},
  \bibinfo{author}{\bibfnamefont{P.}~\bibnamefont{Laguna}}, \bibnamefont{and}
  \bibinfo{author}{\bibfnamefont{R.~A.} \bibnamefont{Matzner}},
  \bibinfo{journal}{Phys. Rev. D} \textbf{\bibinfo{volume}{76}},
  \bibinfo{pages}{084032} (\bibinfo{year}{2007}{\natexlab{c}}),
  \eprint{0706.2541}.

\bibitem[{\citenamefont{Br\"{u}gmann et~al.}(2008)\citenamefont{Br\"{u}gmann,
  Gonzalez, Hannam, Husa, and Sperhake}}]{Brugmann:2007zj}
\bibinfo{author}{\bibfnamefont{B.}~\bibnamefont{Br\"{u}gmann}},
  \bibinfo{author}{\bibfnamefont{J.~A.} \bibnamefont{Gonzalez}},
  \bibinfo{author}{\bibfnamefont{M.}~\bibnamefont{Hannam}},
  \bibinfo{author}{\bibfnamefont{S.}~\bibnamefont{Husa}}, \bibnamefont{and}
  \bibinfo{author}{\bibfnamefont{U.}~\bibnamefont{Sperhake}},
  \bibinfo{journal}{Phys. Rev. D} \textbf{\bibinfo{volume}{77}},
  \bibinfo{pages}{124047} (\bibinfo{year}{2008}), \eprint{0707.0135}.

\bibitem[{\citenamefont{Schnittman et~al.}(2008)}]{Schnittman:2007ij}
\bibinfo{author}{\bibfnamefont{J.~D.} \bibnamefont{Schnittman}}
  \bibnamefont{et~al.}, \bibinfo{journal}{Phys. Rev. D}
  \textbf{\bibinfo{volume}{77}}, \bibinfo{pages}{044031}
  (\bibinfo{year}{2008}), \eprint{0707.0301}.

\bibitem[{\citenamefont{Pollney et~al.}(2007)}]{Pollney:2007ss}
\bibinfo{author}{\bibfnamefont{D.}~\bibnamefont{Pollney}} \bibnamefont{et~al.},
  \bibinfo{journal}{Phys. Rev. D} \textbf{\bibinfo{volume}{76}},
  \bibinfo{pages}{124002} (\bibinfo{year}{2007}), \eprint{0707.2559}.

\bibitem[{\citenamefont{Lousto and Zlochower}(2008)}]{Lousto:2007db}
\bibinfo{author}{\bibfnamefont{C.~O.} \bibnamefont{Lousto}} \bibnamefont{and}
  \bibinfo{author}{\bibfnamefont{Y.}~\bibnamefont{Zlochower}},
  \bibinfo{journal}{Phys. Rev. D} \textbf{\bibinfo{volume}{77}},
  \bibinfo{pages}{044028} (\bibinfo{year}{2008}), \eprint{0708.4048}.

\bibitem[{\citenamefont{Baker et~al.}(2008)}]{Baker:2008md}
\bibinfo{author}{\bibfnamefont{J.~G.} \bibnamefont{Baker}}
  \bibnamefont{et~al.}, \bibinfo{journal}{Astrophys. J.}
  \textbf{\bibinfo{volume}{682}}, \bibinfo{pages}{L29} (\bibinfo{year}{2008}),
  \eprint{0802.0416}.

\bibitem[{\citenamefont{Dain et~al.}(2008)\citenamefont{Dain, Lousto, and
  Zlochower}}]{Dain:2008ck}
\bibinfo{author}{\bibfnamefont{S.}~\bibnamefont{Dain}},
  \bibinfo{author}{\bibfnamefont{C.~O.} \bibnamefont{Lousto}},
  \bibnamefont{and}
  \bibinfo{author}{\bibfnamefont{Y.}~\bibnamefont{Zlochower}},
  \bibinfo{journal}{Phys. Rev. D} \textbf{\bibinfo{volume}{78}},
  \bibinfo{pages}{024039} (\bibinfo{year}{2008}), \eprint{0803.0351}.

\bibitem[{\citenamefont{Healy et~al.}(2009)\citenamefont{Healy, Herrmann,
  Hinder, Shoemaker, Laguna, and Matzner}}]{Healy:2008js}
\bibinfo{author}{\bibfnamefont{J.}~\bibnamefont{Healy}},
  \bibinfo{author}{\bibfnamefont{F.}~\bibnamefont{Herrmann}},
  \bibinfo{author}{\bibfnamefont{I.}~\bibnamefont{Hinder}},
  \bibinfo{author}{\bibfnamefont{D.~M.} \bibnamefont{Shoemaker}},
  \bibinfo{author}{\bibfnamefont{P.}~\bibnamefont{Laguna}}, \bibnamefont{and}
  \bibinfo{author}{\bibfnamefont{R.~A.} \bibnamefont{Matzner}},
  \bibinfo{journal}{Phys. Rev. Lett.} \textbf{\bibinfo{volume}{102}},
  \bibinfo{pages}{041101} (\bibinfo{year}{2009}), \eprint{{arXiv}:0807.3292
  [gr-qc]}.

\bibitem[{\citenamefont{Gonzalez et~al.}(2009)\citenamefont{Gonzalez, Sperhake,
  and Br\"ugmann}}]{Gonzalez:2008bi}
\bibinfo{author}{\bibfnamefont{J.~A.} \bibnamefont{Gonzalez}},
  \bibinfo{author}{\bibfnamefont{U.}~\bibnamefont{Sperhake}}, \bibnamefont{and}
  \bibinfo{author}{\bibfnamefont{B.}~\bibnamefont{Br\"ugmann}},
  \bibinfo{journal}{Phys. Rev. D} \textbf{\bibinfo{volume}{79}},
  \bibinfo{pages}{124006} (\bibinfo{year}{2009}), \eprint{{arXiv}:0811.3952
  [gr-qc]}.

\bibitem[{\citenamefont{Campanelli
  et~al.}(2006{\natexlab{b}})\citenamefont{Campanelli, Lousto, and
  Zlochower}}]{Campanelli:2006fg}
\bibinfo{author}{\bibfnamefont{M.}~\bibnamefont{Campanelli}},
  \bibinfo{author}{\bibfnamefont{C.~O.} \bibnamefont{Lousto}},
  \bibnamefont{and}
  \bibinfo{author}{\bibfnamefont{Y.}~\bibnamefont{Zlochower}},
  \bibinfo{journal}{Phys. Rev. D} \textbf{\bibinfo{volume}{74}},
  \bibinfo{pages}{084023} (\bibinfo{year}{2006}{\natexlab{b}}),
  \eprint{astro-ph/0608275}.

\bibitem[{\citenamefont{Campanelli
  et~al.}(2007{\natexlab{c}})\citenamefont{Campanelli, Lousto, Zlochower,
  Krishnan, and Merritt}}]{Campanelli:2006fy}
\bibinfo{author}{\bibfnamefont{M.}~\bibnamefont{Campanelli}},
  \bibinfo{author}{\bibfnamefont{C.~O.} \bibnamefont{Lousto}},
  \bibinfo{author}{\bibfnamefont{Y.}~\bibnamefont{Zlochower}},
  \bibinfo{author}{\bibfnamefont{B.}~\bibnamefont{Krishnan}}, \bibnamefont{and}
  \bibinfo{author}{\bibfnamefont{D.}~\bibnamefont{Merritt}},
  \bibinfo{journal}{Phys. Rev. D} \textbf{\bibinfo{volume}{75}},
  \bibinfo{pages}{064030} (\bibinfo{year}{2007}{\natexlab{c}}),
  \eprint{gr-qc/0612076}.

\bibitem[{\citenamefont{Berti et~al.}(2007)}]{Berti:2007fi}
\bibinfo{author}{\bibfnamefont{E.}~\bibnamefont{Berti}} \bibnamefont{et~al.},
  \bibinfo{journal}{Phys. Rev. D} \textbf{\bibinfo{volume}{76}},
  \bibinfo{pages}{064034} (\bibinfo{year}{2007}), \eprint{gr-qc/0703053}.

\bibitem[{\citenamefont{Rezzolla et~al.}(2008{\natexlab{a}})}]{Rezzolla:2007xa}
\bibinfo{author}{\bibfnamefont{L.}~\bibnamefont{Rezzolla}}
  \bibnamefont{et~al.}, \bibinfo{journal}{Astrophys. J.}
  \textbf{\bibinfo{volume}{679}}, \bibinfo{pages}{1422}
  (\bibinfo{year}{2008}{\natexlab{a}}), \eprint{0708.3999}.

\bibitem[{\citenamefont{Boyle et~al.}(2008{\natexlab{a}})\citenamefont{Boyle,
  Kesden, and Nissanke}}]{Boyle:2007sz}
\bibinfo{author}{\bibfnamefont{L.}~\bibnamefont{Boyle}},
  \bibinfo{author}{\bibfnamefont{M.}~\bibnamefont{Kesden}}, \bibnamefont{and}
  \bibinfo{author}{\bibfnamefont{S.}~\bibnamefont{Nissanke}},
  \bibinfo{journal}{Phys. Rev. Lett.} \textbf{\bibinfo{volume}{100}},
  \bibinfo{pages}{151101} (\bibinfo{year}{2008}{\natexlab{a}}),
  \eprint{0709.0299}.

\bibitem[{\citenamefont{Rezzolla et~al.}(2008{\natexlab{b}})}]{Rezzolla:2007rd}
\bibinfo{author}{\bibfnamefont{L.}~\bibnamefont{Rezzolla}}
  \bibnamefont{et~al.}, \bibinfo{journal}{Astrophys. J.}
  \textbf{\bibinfo{volume}{674}}, \bibinfo{pages}{L29}
  (\bibinfo{year}{2008}{\natexlab{b}}), \eprint{0710.3345}.

\bibitem[{\citenamefont{Marronetti et~al.}(2008)\citenamefont{Marronetti,
  Tichy, Br\"{u}gmann, Gonzalez, and Sperhake}}]{Marronetti:2007wz}
\bibinfo{author}{\bibfnamefont{P.}~\bibnamefont{Marronetti}},
  \bibinfo{author}{\bibfnamefont{W.}~\bibnamefont{Tichy}},
  \bibinfo{author}{\bibfnamefont{B.}~\bibnamefont{Br\"{u}gmann}},
  \bibinfo{author}{\bibfnamefont{J.}~\bibnamefont{Gonzalez}}, \bibnamefont{and}
  \bibinfo{author}{\bibfnamefont{U.}~\bibnamefont{Sperhake}},
  \bibinfo{journal}{Phys. Rev. D} \textbf{\bibinfo{volume}{77}},
  \bibinfo{pages}{064010} (\bibinfo{year}{2008}), \eprint{0709.2160}.

\bibitem[{\citenamefont{Sperhake et~al.}(2008)}]{Sperhake:2007gu}
\bibinfo{author}{\bibfnamefont{U.}~\bibnamefont{Sperhake}}
  \bibnamefont{et~al.}, \bibinfo{journal}{Phys. Rev. D}
  \textbf{\bibinfo{volume}{78}}, \bibinfo{pages}{064069}
  (\bibinfo{year}{2008}), \eprint{0710.3823}.

\bibitem[{\citenamefont{Hinder et~al.}(2008)\citenamefont{Hinder, Vaishnav,
  Herrmann, Shoemaker, and Laguna}}]{Hinder:2007qu}
\bibinfo{author}{\bibfnamefont{I.}~\bibnamefont{Hinder}},
  \bibinfo{author}{\bibfnamefont{B.}~\bibnamefont{Vaishnav}},
  \bibinfo{author}{\bibfnamefont{F.}~\bibnamefont{Herrmann}},
  \bibinfo{author}{\bibfnamefont{D.}~\bibnamefont{Shoemaker}},
  \bibnamefont{and} \bibinfo{author}{\bibfnamefont{P.}~\bibnamefont{Laguna}},
  \bibinfo{journal}{Phys. Rev. D} \textbf{\bibinfo{volume}{77}},
  \bibinfo{pages}{081502} (\bibinfo{year}{2008}), \eprint{0710.5167}.

\bibitem[{\citenamefont{Berti et~al.}(2008)\citenamefont{Berti, Cardoso,
  Gonzalez, Sperhake, and Br\"{u}gmann}}]{Berti:2007nw}
\bibinfo{author}{\bibfnamefont{E.}~\bibnamefont{Berti}},
  \bibinfo{author}{\bibfnamefont{V.}~\bibnamefont{Cardoso}},
  \bibinfo{author}{\bibfnamefont{J.~A.} \bibnamefont{Gonzalez}},
  \bibinfo{author}{\bibfnamefont{U.}~\bibnamefont{Sperhake}}, \bibnamefont{and}
  \bibinfo{author}{\bibfnamefont{B.}~\bibnamefont{Br\"{u}gmann}},
  \bibinfo{journal}{Class. Quantum Grav.} \textbf{\bibinfo{volume}{25}},
  \bibinfo{pages}{114035} (\bibinfo{year}{2008}), \eprint{0711.1097}.

\bibitem[{\citenamefont{Boyle and Kesden}(2008)}]{Boyle:2007ru}
\bibinfo{author}{\bibfnamefont{L.}~\bibnamefont{Boyle}} \bibnamefont{and}
  \bibinfo{author}{\bibfnamefont{M.}~\bibnamefont{Kesden}},
  \bibinfo{journal}{Phys. Rev. D} \textbf{\bibinfo{volume}{78}},
  \bibinfo{pages}{024017} (\bibinfo{year}{2008}), \eprint{0712.2819}.

\bibitem[{\citenamefont{Tichy and Marronetti}(2008)}]{Tichy:2008du}
\bibinfo{author}{\bibfnamefont{W.}~\bibnamefont{Tichy}} \bibnamefont{and}
  \bibinfo{author}{\bibfnamefont{P.}~\bibnamefont{Marronetti}},
  \bibinfo{journal}{Phys. Rev. D} \textbf{\bibinfo{volume}{78}},
  \bibinfo{pages}{081501} (\bibinfo{year}{2008}), \eprint{0807.2985}.

\bibitem[{\citenamefont{Rezzolla}(2009)}]{Rezzolla:2008sd}
\bibinfo{author}{\bibfnamefont{L.}~\bibnamefont{Rezzolla}},
  \bibinfo{journal}{Class. Quant. Grav.} \textbf{\bibinfo{volume}{26}},
  \bibinfo{pages}{094023} (\bibinfo{year}{2009}), \bibinfo{note}{proceedings of
  7th LISA Symposium, Barcelona June 2008}, \eprint{{arXiv}:0812.2325 [gr-qc]}.

\bibitem[{\citenamefont{Baker et~al.}(2007{\natexlab{b}})\citenamefont{Baker,
  van Meter, McWilliams, Centrella, and Kelly}}]{Baker:2006ha}
\bibinfo{author}{\bibfnamefont{J.~G.} \bibnamefont{Baker}},
  \bibinfo{author}{\bibfnamefont{J.~R.} \bibnamefont{van Meter}},
  \bibinfo{author}{\bibfnamefont{S.~T.} \bibnamefont{McWilliams}},
  \bibinfo{author}{\bibfnamefont{J.}~\bibnamefont{Centrella}},
  \bibnamefont{and} \bibinfo{author}{\bibfnamefont{B.~J.} \bibnamefont{Kelly}},
  \bibinfo{journal}{Phys. Rev. Lett.} \textbf{\bibinfo{volume}{99}},
  \bibinfo{pages}{181101} (\bibinfo{year}{2007}{\natexlab{b}}),
  \eprint{gr-qc/0612024}.

\bibitem[{\citenamefont{Hannam et~al.}(2008{\natexlab{a}})}]{Hannam:2007ik}
\bibinfo{author}{\bibfnamefont{M.}~\bibnamefont{Hannam}} \bibnamefont{et~al.},
  \bibinfo{journal}{Phys.~Rev.~D} \textbf{\bibinfo{volume}{77}},
  \bibinfo{pages}{044020} (\bibinfo{year}{2008}{\natexlab{a}}),
  \eprint{arXiv:0706.1305}.

\bibitem[{\citenamefont{Boyle et~al.}(2007)}]{Boyle:2007ft}
\bibinfo{author}{\bibfnamefont{M.}~\bibnamefont{Boyle}} \bibnamefont{et~al.},
  \bibinfo{journal}{Phys.~Rev.~D} \textbf{\bibinfo{volume}{76}},
  \bibinfo{pages}{124038} (\bibinfo{year}{2007}), \eprint{arXiv:0710.0158}.

\bibitem[{\citenamefont{Hannam et~al.}(2008{\natexlab{b}})\citenamefont{Hannam,
  Husa, Br\"{u}gmann, and Gopakumar}}]{Hannam:2007wf}
\bibinfo{author}{\bibfnamefont{M.}~\bibnamefont{Hannam}},
  \bibinfo{author}{\bibfnamefont{S.}~\bibnamefont{Husa}},
  \bibinfo{author}{\bibfnamefont{B.}~\bibnamefont{Br\"{u}gmann}},
  \bibnamefont{and}
  \bibinfo{author}{\bibfnamefont{A.}~\bibnamefont{Gopakumar}},
  \bibinfo{journal}{Phys. Rev. D} \textbf{\bibinfo{volume}{78}},
  \bibinfo{pages}{104007} (\bibinfo{year}{2008}{\natexlab{b}}),
  \eprint{0712.3787}.

\bibitem[{\citenamefont{Campanelli et~al.}(2009)\citenamefont{Campanelli,
  Lousto, Nakano, and Zlochower}}]{Campanelli:2008nk}
\bibinfo{author}{\bibfnamefont{M.}~\bibnamefont{Campanelli}},
  \bibinfo{author}{\bibfnamefont{C.~O.} \bibnamefont{Lousto}},
  \bibinfo{author}{\bibfnamefont{H.}~\bibnamefont{Nakano}}, \bibnamefont{and}
  \bibinfo{author}{\bibfnamefont{Y.}~\bibnamefont{Zlochower}},
  \bibinfo{journal}{Phys. Rev. D} \textbf{\bibinfo{volume}{79}},
  \bibinfo{pages}{084010} (\bibinfo{year}{2009}), \eprint{0808.0713}.

\bibitem[{\citenamefont{Boyle et~al.}(2008{\natexlab{b}})}]{Boyle:2008ge}
\bibinfo{author}{\bibfnamefont{M.}~\bibnamefont{Boyle}} \bibnamefont{et~al.},
  \bibinfo{journal}{Phys. Rev. D} \textbf{\bibinfo{volume}{78}},
  \bibinfo{pages}{104020} (\bibinfo{year}{2008}{\natexlab{b}}),
  \eprint{0804.4184}.

\bibitem[{\citenamefont{Hinder et~al.}(2010)\citenamefont{Hinder, Herrmann,
  Laguna, and Shoemaker}}]{hinder-2008}
\bibinfo{author}{\bibfnamefont{I.}~\bibnamefont{Hinder}},
  \bibinfo{author}{\bibfnamefont{F.}~\bibnamefont{Herrmann}},
  \bibinfo{author}{\bibfnamefont{P.}~\bibnamefont{Laguna}}, \bibnamefont{and}
  \bibinfo{author}{\bibfnamefont{D.}~\bibnamefont{Shoemaker}},
  \bibinfo{journal}{Phys. Rev.} \textbf{\bibinfo{volume}{D82}},
  \bibinfo{pages}{024033} (\bibinfo{year}{2010}), \eprint{0806.1037}.

\bibitem[{\citenamefont{Chu et~al.}(2009)\citenamefont{Chu, Pfeiffer, and
  Scheel}}]{chu-2009}
\bibinfo{author}{\bibfnamefont{T.}~\bibnamefont{Chu}},
  \bibinfo{author}{\bibfnamefont{H.~P.} \bibnamefont{Pfeiffer}},
  \bibnamefont{and} \bibinfo{author}{\bibfnamefont{M.~A.}
  \bibnamefont{Scheel}}, \bibinfo{journal}{Phys. Rev.}
  \textbf{\bibinfo{volume}{D80}}, \bibinfo{pages}{124051}
  (\bibinfo{year}{2009}), \eprint{0909.1313}.

\bibitem[{\citenamefont{Pollney et~al.}(2009)\citenamefont{Pollney, Reisswig,
  Schnetter, Dorband, and Diener}}]{pollney-2009}
\bibinfo{author}{\bibfnamefont{D.}~\bibnamefont{Pollney}},
  \bibinfo{author}{\bibfnamefont{C.}~\bibnamefont{Reisswig}},
  \bibinfo{author}{\bibfnamefont{E.}~\bibnamefont{Schnetter}},
  \bibinfo{author}{\bibfnamefont{N.}~\bibnamefont{Dorband}}, \bibnamefont{and}
  \bibinfo{author}{\bibfnamefont{P.}~\bibnamefont{Diener}}
  (\bibinfo{year}{2009}), \eprint{0910.3803}.

\bibitem[{\citenamefont{Buonanno
  et~al.}(2007{\natexlab{a}})\citenamefont{Buonanno, Cook, and
  Pretorius}}]{Buonanno:2006ui}
\bibinfo{author}{\bibfnamefont{A.}~\bibnamefont{Buonanno}},
  \bibinfo{author}{\bibfnamefont{G.~B.} \bibnamefont{Cook}}, \bibnamefont{and}
  \bibinfo{author}{\bibfnamefont{F.}~\bibnamefont{Pretorius}},
  \bibinfo{journal}{Phys.~Rev.~D} \textbf{\bibinfo{volume}{75}},
  \bibinfo{pages}{124018} (\bibinfo{year}{2007}{\natexlab{a}}).

\bibitem[{\citenamefont{Pan et~al.}(2008)\citenamefont{Pan, Buonanno, Baker,
  Centrella, Kelly, McWilliams, Pretorius, and van Meter}}]{Pan2007}
\bibinfo{author}{\bibfnamefont{Y.}~\bibnamefont{Pan}},
  \bibinfo{author}{\bibfnamefont{A.}~\bibnamefont{Buonanno}},
  \bibinfo{author}{\bibfnamefont{J.~G.} \bibnamefont{Baker}},
  \bibinfo{author}{\bibfnamefont{J.}~\bibnamefont{Centrella}},
  \bibinfo{author}{\bibfnamefont{B.~J.} \bibnamefont{Kelly}},
  \bibinfo{author}{\bibfnamefont{S.~T.} \bibnamefont{McWilliams}},
  \bibinfo{author}{\bibfnamefont{F.}~\bibnamefont{Pretorius}},
  \bibnamefont{and} \bibinfo{author}{\bibfnamefont{J.~R.} \bibnamefont{van
  Meter}}, \bibinfo{journal}{Phys. Rev. D} \textbf{\bibinfo{volume}{77}},
  \bibinfo{pages}{024014} (\bibinfo{year}{2008}).

\bibitem[{\citenamefont{Buonanno et~al.}(2007{\natexlab{b}})}]{Buonanno:2007pf}
\bibinfo{author}{\bibfnamefont{A.}~\bibnamefont{Buonanno}}
  \bibnamefont{et~al.}, \bibinfo{journal}{Phys.~Rev.~D}
  \textbf{\bibinfo{volume}{76}}, \bibinfo{pages}{104049}
  (\bibinfo{year}{2007}{\natexlab{b}}), \eprint{0706.3732}.

\bibitem[{\citenamefont{Damour et~al.}(2008)\citenamefont{Damour, Nagar,
  Dorband, Pollney, and Rezzolla}}]{DN2007b}
\bibinfo{author}{\bibfnamefont{T.}~\bibnamefont{Damour}},
  \bibinfo{author}{\bibfnamefont{A.}~\bibnamefont{Nagar}},
  \bibinfo{author}{\bibfnamefont{E.~N.} \bibnamefont{Dorband}},
  \bibinfo{author}{\bibfnamefont{D.}~\bibnamefont{Pollney}}, \bibnamefont{and}
  \bibinfo{author}{\bibfnamefont{L.}~\bibnamefont{Rezzolla}},
  \bibinfo{journal}{Phys. Rev. D} \textbf{\bibinfo{volume}{77}},
  \bibinfo{eid}{084017} (\bibinfo{year}{2008}).

\bibitem[{\citenamefont{{Damour} et~al.}(2008)\citenamefont{{Damour}, {Nagar},
  {Hannam}, {Husa}, and {Br{\"u}gmann}}}]{DN2008}
\bibinfo{author}{\bibfnamefont{T.}~\bibnamefont{{Damour}}},
  \bibinfo{author}{\bibfnamefont{A.}~\bibnamefont{{Nagar}}},
  \bibinfo{author}{\bibfnamefont{M.}~\bibnamefont{{Hannam}}},
  \bibinfo{author}{\bibfnamefont{S.}~\bibnamefont{{Husa}}}, \bibnamefont{and}
  \bibinfo{author}{\bibfnamefont{B.}~\bibnamefont{{Br{\"u}gmann}}},
  \bibinfo{journal}{Phys. Rev. D} \textbf{\bibinfo{volume}{78}},
  \bibinfo{pages}{044039} (\bibinfo{year}{2008}).

\bibitem[{\citenamefont{Boyle et~al.}(2008{\natexlab{c}})\citenamefont{Boyle,
  Buonanno, Kidder, Mrou\'e, Pan, Pfeiffer, and Scheel}}]{Boyle2008a}
\bibinfo{author}{\bibfnamefont{M.}~\bibnamefont{Boyle}},
  \bibinfo{author}{\bibfnamefont{A.}~\bibnamefont{Buonanno}},
  \bibinfo{author}{\bibfnamefont{L.~E.} \bibnamefont{Kidder}},
  \bibinfo{author}{\bibfnamefont{A.~H.} \bibnamefont{Mrou\'e}},
  \bibinfo{author}{\bibfnamefont{Y.}~\bibnamefont{Pan}},
  \bibinfo{author}{\bibfnamefont{H.~P.} \bibnamefont{Pfeiffer}},
  \bibnamefont{and} \bibinfo{author}{\bibfnamefont{M.~A.}
  \bibnamefont{Scheel}}, \textbf{\bibinfo{volume}{78}}, \bibinfo{pages}{104020}
  (\bibinfo{year}{2008}{\natexlab{c}}).

\bibitem[{\citenamefont{Damour and Nagar}(2009)}]{Damour2009a}
\bibinfo{author}{\bibfnamefont{T.}~\bibnamefont{Damour}} \bibnamefont{and}
  \bibinfo{author}{\bibfnamefont{A.}~\bibnamefont{Nagar}},
  \bibinfo{journal}{Phys. Rev. D} \textbf{\bibinfo{volume}{79}},
  \bibinfo{pages}{081503} (\bibinfo{year}{2009}).

\bibitem[{\citenamefont{Buonanno et~al.}(2009)\citenamefont{Buonanno, Pan,
  Pfeiffer, Scheel, Buchman, and Kidder}}]{Buonanno:2009qa}
\bibinfo{author}{\bibfnamefont{A.}~\bibnamefont{Buonanno}},
  \bibinfo{author}{\bibfnamefont{Y.}~\bibnamefont{Pan}},
  \bibinfo{author}{\bibfnamefont{H.~P.} \bibnamefont{Pfeiffer}},
  \bibinfo{author}{\bibfnamefont{M.~A.} \bibnamefont{Scheel}},
  \bibinfo{author}{\bibfnamefont{L.~T.} \bibnamefont{Buchman}},
  \bibnamefont{and} \bibinfo{author}{\bibfnamefont{L.~E.}
  \bibnamefont{Kidder}}, \bibinfo{journal}{\prd} \textbf{\bibinfo{volume}{79}},
  \bibinfo{pages}{124028} (\bibinfo{year}{2009}).

\bibitem[{\citenamefont{Ajith et~al.}(2007)}]{Ajith:2007qp}
\bibinfo{author}{\bibfnamefont{P.}~\bibnamefont{Ajith}} \bibnamefont{et~al.},
  \bibinfo{journal}{Class. Quantum Grav.} \textbf{\bibinfo{volume}{24}},
  \bibinfo{pages}{S689} (\bibinfo{year}{2007}), \eprint{arXiv:0704.3764}.

\bibitem[{\citenamefont{{Ajith} et~al.}(2009)\citenamefont{{Ajith}, {Hannam},
  {Husa}, {Chen}, {Bruegmann}, {Dorband}, {Mueller}, {Ohme}, {Pollney},
  {Reisswig} et~al.}}]{2009arXiv0909.2867A}
\bibinfo{author}{\bibfnamefont{P.}~\bibnamefont{{Ajith}}},
  \bibinfo{author}{\bibfnamefont{M.}~\bibnamefont{{Hannam}}},
  \bibinfo{author}{\bibfnamefont{S.}~\bibnamefont{{Husa}}},
  \bibinfo{author}{\bibfnamefont{Y.}~\bibnamefont{{Chen}}},
  \bibinfo{author}{\bibfnamefont{B.}~\bibnamefont{{Bruegmann}}},
  \bibinfo{author}{\bibfnamefont{N.}~\bibnamefont{{Dorband}}},
  \bibinfo{author}{\bibfnamefont{D.}~\bibnamefont{{Mueller}}},
  \bibinfo{author}{\bibfnamefont{F.}~\bibnamefont{{Ohme}}},
  \bibinfo{author}{\bibfnamefont{D.}~\bibnamefont{{Pollney}}},
  \bibinfo{author}{\bibfnamefont{C.}~\bibnamefont{{Reisswig}}},
  \bibnamefont{et~al.}, \bibinfo{journal}{ArXiv e-prints}
  (\bibinfo{year}{2009}), \eprint{0909.2867}.

\bibitem[{\citenamefont{Abbott et~al.}(2006)}]{LIGOS2bbh}
\bibinfo{author}{\bibfnamefont{B.}~\bibnamefont{Abbott}} \bibnamefont{et~al.}
  (\bibinfo{collaboration}{{LIGO} Scientific Collaboration}),
  \bibinfo{journal}{Phys.~Rev.~D} \textbf{\bibinfo{volume}{73}},
  \bibinfo{pages}{062001} (\bibinfo{year}{2006}), \eprint{arXiv:gr-qc/0509129}.

\bibitem[{\citenamefont{Abbott et~al.}(2008{\natexlab{a}})}]{LIGOS3S4all}
\bibinfo{author}{\bibfnamefont{B.}~\bibnamefont{Abbott}} \bibnamefont{et~al.}
  (\bibinfo{collaboration}{{LIGO} Scientific Collaboration}),
  \bibinfo{journal}{Phys.~Rev.~D} \textbf{\bibinfo{volume}{77}},
  \bibinfo{pages}{062002} (\bibinfo{year}{2008}{\natexlab{a}}),
  \eprint{arXiv:0704.3368}.

\bibitem[{\citenamefont{Kopparapu et~al.}(2008)\citenamefont{Kopparapu, Hanna,
  Kalogera, O'Shaughnessy, Gonzalez, Brady, and Fairhurst}}]{LIGOS3S4Galaxies}
\bibinfo{author}{\bibfnamefont{R.~K.} \bibnamefont{Kopparapu}},
  \bibinfo{author}{\bibfnamefont{C.}~\bibnamefont{Hanna}},
  \bibinfo{author}{\bibfnamefont{V.}~\bibnamefont{Kalogera}},
  \bibinfo{author}{\bibfnamefont{R.}~\bibnamefont{O'Shaughnessy}},
  \bibinfo{author}{\bibfnamefont{G.}~\bibnamefont{Gonzalez}},
  \bibinfo{author}{\bibfnamefont{P.~R.} \bibnamefont{Brady}}, \bibnamefont{and}
  \bibinfo{author}{\bibfnamefont{S.}~\bibnamefont{Fairhurst}},
  \bibinfo{journal}{\apj} \textbf{\bibinfo{volume}{675}}, \bibinfo{pages}{1459}
  (\bibinfo{year}{2008}).

\bibitem[{\citenamefont{Buonanno et~al.}(2003)\citenamefont{Buonanno, Chen, and
  Vallisneri}}]{BuonannoChenVallisneri:2003a}
\bibinfo{author}{\bibfnamefont{A.}~\bibnamefont{Buonanno}},
  \bibinfo{author}{\bibfnamefont{Y.}~\bibnamefont{Chen}}, \bibnamefont{and}
  \bibinfo{author}{\bibfnamefont{M.}~\bibnamefont{Vallisneri}},
  \bibinfo{journal}{Phys.~Rev.~D} \textbf{\bibinfo{volume}{67}},
  \bibinfo{pages}{024016} (\bibinfo{year}{2003}), \bibinfo{note}{erratum-ibid.
  74 (2006) 029903(E)}.

\bibitem[{\citenamefont{Abbott et~al.}(2008{\natexlab{b}})}]{S3_BCVSpin}
\bibinfo{author}{\bibfnamefont{B.}~\bibnamefont{Abbott}} \bibnamefont{et~al.}
  (\bibinfo{collaboration}{{LIGO} Scientific Collaboration}),
  \bibinfo{journal}{Phys.~Rev.~D} \textbf{\bibinfo{volume}{78}},
  \bibinfo{pages}{042002} (\bibinfo{year}{2008}{\natexlab{b}}),
  \eprint{arXiv:0712.2050}.

\bibitem[{\citenamefont{Abbott et~al.}(2009{\natexlab{d}})}]{Abbott:2009km}
\bibinfo{author}{\bibfnamefont{B.}~\bibnamefont{Abbott}} \bibnamefont{et~al.}
  (\bibinfo{collaboration}{LIGO Scientific Collaboration}),
  \bibinfo{journal}{Phys.~Rev.~D} \textbf{\bibinfo{volume}{80}},
  \bibinfo{pages}{062001} (\bibinfo{year}{2009}{\natexlab{d}}),
  \eprint{0905.1654}.

\bibitem[{\citenamefont{Cutler et~al.}(1993)}]{Cutler:1992tc}
\bibinfo{author}{\bibfnamefont{C.}~\bibnamefont{Cutler}} \bibnamefont{et~al.},
  \bibinfo{journal}{Phys. Rev. Lett.} \textbf{\bibinfo{volume}{70}},
  \bibinfo{pages}{2984} (\bibinfo{year}{1993}).

\bibitem[{\citenamefont{McKechan et~al.}(2010)\citenamefont{McKechan, Robinson,
  and Sathyaprakash}}]{MRS:2010}
\bibinfo{author}{\bibfnamefont{D.~J.~A.} \bibnamefont{McKechan}},
  \bibinfo{author}{\bibfnamefont{C.}~\bibnamefont{Robinson}}, \bibnamefont{and}
  \bibinfo{author}{\bibfnamefont{B.~S.} \bibnamefont{Sathyaprakash}},
  \bibinfo{journal}{Class. Quantum Grav.} \textbf{\bibinfo{volume}{27}},
  \bibinfo{pages}{084020} (\bibinfo{year}{2010}).

\bibitem[{\citenamefont{Buonanno and Damour}(1999)}]{BuonannoDamour:1999}
\bibinfo{author}{\bibfnamefont{A.}~\bibnamefont{Buonanno}} \bibnamefont{and}
  \bibinfo{author}{\bibfnamefont{T.}~\bibnamefont{Damour}},
  \bibinfo{journal}{Phys.~Rev.~D} \textbf{\bibinfo{volume}{59}},
  \bibinfo{pages}{084006} (\bibinfo{year}{1999}).

\bibitem[{\citenamefont{Buonanno and Damour}(2000)}]{BuonannoDamour:2000}
\bibinfo{author}{\bibfnamefont{A.}~\bibnamefont{Buonanno}} \bibnamefont{and}
  \bibinfo{author}{\bibfnamefont{T.}~\bibnamefont{Damour}},
  \bibinfo{journal}{Phys.~Rev.~D} \textbf{\bibinfo{volume}{62}},
  \bibinfo{pages}{064015} (\bibinfo{year}{2000}).

\bibitem[{\citenamefont{Damour et~al.}(2000)\citenamefont{Damour, Jaranowski,
  and Sch{\"a}fer}}]{DamourJaranowskiSchaefer:2000}
\bibinfo{author}{\bibfnamefont{T.}~\bibnamefont{Damour}},
  \bibinfo{author}{\bibfnamefont{P.}~\bibnamefont{Jaranowski}},
  \bibnamefont{and}
  \bibinfo{author}{\bibfnamefont{G.}~\bibnamefont{Sch{\"a}fer}},
  \bibinfo{journal}{Phys.~Rev.~D} \textbf{\bibinfo{volume}{62}},
  \bibinfo{pages}{084011} (\bibinfo{year}{2000}).

\bibitem[{\citenamefont{Damour et~al.}(2003)\citenamefont{Damour, Iyer,
  Jaranowski, and Sathyaprakash}}]{Damour03}
\bibinfo{author}{\bibfnamefont{T.}~\bibnamefont{Damour}},
  \bibinfo{author}{\bibfnamefont{B.~R.} \bibnamefont{Iyer}},
  \bibinfo{author}{\bibfnamefont{P.}~\bibnamefont{Jaranowski}},
  \bibnamefont{and} \bibinfo{author}{\bibfnamefont{B.~S.}
  \bibnamefont{Sathyaprakash}}, \bibinfo{journal}{Phys. Rev. D}
  \textbf{\bibinfo{volume}{67}}, \bibinfo{pages}{064028}
  (\bibinfo{year}{2003}).

\bibitem[{\citenamefont{Damour et~al.}(1998)\citenamefont{Damour, Iyer, and
  Sathyaprakash}}]{Damour:1998zb}
\bibinfo{author}{\bibfnamefont{T.}~\bibnamefont{Damour}},
  \bibinfo{author}{\bibfnamefont{B.~R.} \bibnamefont{Iyer}}, \bibnamefont{and}
  \bibinfo{author}{\bibfnamefont{B.~S.} \bibnamefont{Sathyaprakash}},
  \bibinfo{journal}{Phys.~Rev.~D} \textbf{\bibinfo{volume}{57}},
  \bibinfo{pages}{885} (\bibinfo{year}{1998}).

\bibitem[{\citenamefont{Damour et~al.}(2009)\citenamefont{Damour, Iyer, and
  Nagar}}]{DIN}
\bibinfo{author}{\bibfnamefont{T.}~\bibnamefont{Damour}},
  \bibinfo{author}{\bibfnamefont{B.~R.} \bibnamefont{Iyer}}, \bibnamefont{and}
  \bibinfo{author}{\bibfnamefont{A.}~\bibnamefont{Nagar}},
  \bibinfo{journal}{Phys. Rev. D} \textbf{\bibinfo{volume}{79}},
  \bibinfo{pages}{064004} (\bibinfo{year}{2009}).

\bibitem[{\citenamefont{Pan et~al.}(2011)\citenamefont{Pan, Buonanno, Boyle,
  Buchman, Kidder, Pfeiffer, and Scheel}}]{Pan:2011}
\bibinfo{author}{\bibfnamefont{Y.}~\bibnamefont{Pan}},
  \bibinfo{author}{\bibfnamefont{A.}~\bibnamefont{Buonanno}},
  \bibinfo{author}{\bibfnamefont{M.}~\bibnamefont{Boyle}},
  \bibinfo{author}{\bibfnamefont{L.}~\bibnamefont{Buchman}},
  \bibinfo{author}{\bibfnamefont{L.}~\bibnamefont{Kidder}},
  \bibinfo{author}{\bibfnamefont{H.~P.} \bibnamefont{Pfeiffer}},
  \bibnamefont{and} \bibinfo{author}{\bibfnamefont{M.}~\bibnamefont{Scheel}}
  (\bibinfo{year}{2011}), \bibinfo{note}{in preparation}.

\bibitem[{\citenamefont{Damour and Gopakumar}(2006)}]{Damour06}
\bibinfo{author}{\bibfnamefont{T.}~\bibnamefont{Damour}} \bibnamefont{and}
  \bibinfo{author}{\bibfnamefont{A.}~\bibnamefont{Gopakumar}},
  \bibinfo{journal}{Phys. Rev. D} \textbf{\bibinfo{volume}{73}},
  \bibinfo{pages}{124006} (\bibinfo{year}{2006}).

\bibitem[{\citenamefont{{Berti} et~al.}(2006)\citenamefont{{Berti}, {Cardoso},
  and {Will}}}]{2006PhRvD..73f4030B}
\bibinfo{author}{\bibfnamefont{E.}~\bibnamefont{{Berti}}},
  \bibinfo{author}{\bibfnamefont{V.}~\bibnamefont{{Cardoso}}},
  \bibnamefont{and} \bibinfo{author}{\bibfnamefont{C.~M.}
  \bibnamefont{{Will}}}, \bibinfo{journal}{Phys.~Rev.~D}
  \textbf{\bibinfo{volume}{73}}, \bibinfo{pages}{064030}
  (\bibinfo{year}{2006}).

\bibitem[{\citenamefont{Ajith et~al.}(2008)}]{Ajith:2007kx}
\bibinfo{author}{\bibfnamefont{P.}~\bibnamefont{Ajith}} \bibnamefont{et~al.},
  \bibinfo{journal}{Phys.~Rev.~D} \textbf{\bibinfo{volume}{77}},
  \bibinfo{pages}{104017} (\bibinfo{year}{2008}), \eprint{arXiv:0710.2335}.

\bibitem[{\citenamefont{Bruegmann et~al.}(2008)\citenamefont{Bruegmann,
  Gonzalez, Hannam, Husa, Sperhake, and Tichy}}]{Bruegmann:2008}
\bibinfo{author}{\bibfnamefont{B.}~\bibnamefont{Bruegmann}},
  \bibinfo{author}{\bibfnamefont{J.~A.} \bibnamefont{Gonzalez}},
  \bibinfo{author}{\bibfnamefont{M.}~\bibnamefont{Hannam}},
  \bibinfo{author}{\bibfnamefont{S.}~\bibnamefont{Husa}},
  \bibinfo{author}{\bibfnamefont{U.}~\bibnamefont{Sperhake}}, \bibnamefont{and}
  \bibinfo{author}{\bibfnamefont{W.}~\bibnamefont{Tichy}},
  \bibinfo{journal}{\prd} \textbf{\bibinfo{volume}{77}},
  \bibinfo{pages}{024027} (\bibinfo{year}{2008}), \eprint{gr-qc/0610128}.

\bibitem[{\citenamefont{Ajith}(2008)}]{Ajith:2007xh}
\bibinfo{author}{\bibfnamefont{P.}~\bibnamefont{Ajith}},
  \bibinfo{journal}{Class. Quantum Grav.} \textbf{\bibinfo{volume}{25}},
  \bibinfo{pages}{114033} (\bibinfo{year}{2008}), \eprint{arXiv:0712.0343}.

\bibitem[{\citenamefont{Allen}(2005)}]{Allen:2004}
\bibinfo{author}{\bibfnamefont{B.}~\bibnamefont{Allen}},
  \bibinfo{journal}{Phys.~Rev.~D} \textbf{\bibinfo{volume}{71}},
  \bibinfo{pages}{062001} (\bibinfo{year}{2005}).

\bibitem[{\citenamefont{Cokelaer}(2007)}]{hexabank}
\bibinfo{author}{\bibfnamefont{T.}~\bibnamefont{Cokelaer}},
  \bibinfo{journal}{Phys.~Rev.~D} \textbf{\bibinfo{volume}{76}},
  \bibinfo{pages}{102004} (\bibinfo{year}{2007}), \eprint{arXiv:0706.4437}.

\bibitem[{\citenamefont{Owen}(1996)}]{Owen:1995tm}
\bibinfo{author}{\bibfnamefont{B.~J.} \bibnamefont{Owen}},
  \bibinfo{journal}{Phys.~Rev.~D} \textbf{\bibinfo{volume}{53}},
  \bibinfo{pages}{6749} (\bibinfo{year}{1996}).

\bibitem[{\citenamefont{Babak et~al.}(2006)\citenamefont{Babak,
  Balasubramanian, Churches, Cokelaer, and Sathyaprakash}}]{BBCCS:2006}
\bibinfo{author}{\bibfnamefont{S.}~\bibnamefont{Babak}},
  \bibinfo{author}{\bibnamefont{Balasubramanian}},
  \bibinfo{author}{\bibfnamefont{D.}~\bibnamefont{Churches}},
  \bibinfo{author}{\bibfnamefont{T.}~\bibnamefont{Cokelaer}}, \bibnamefont{and}
  \bibinfo{author}{\bibfnamefont{B.}~\bibnamefont{Sathyaprakash}},
  \bibinfo{journal}{Class. Quantum Grav.} \textbf{\bibinfo{volume}{23}},
  \bibinfo{pages}{5477} (\bibinfo{year}{2006}), \eprint{gr-qc/0604037}.

\bibitem[{\citenamefont{Robinson et~al.}(2008)\citenamefont{Robinson,
  Sathyaprakash, and Sengupta}}]{Robinson:2008}
\bibinfo{author}{\bibfnamefont{C.~A.~K.} \bibnamefont{Robinson}},
  \bibinfo{author}{\bibfnamefont{B.~S.} \bibnamefont{Sathyaprakash}},
  \bibnamefont{and} \bibinfo{author}{\bibfnamefont{A.~S.}
  \bibnamefont{Sengupta}}, \bibinfo{journal}{Phys.~Rev.~D}
  \textbf{\bibinfo{volume}{78}}, \bibinfo{eid}{062002} (\bibinfo{year}{2008}).

\bibitem[{\citenamefont{Slutsky et~al.}(2010)}]{Slutsky:2010ff}
\bibinfo{author}{\bibfnamefont{J.}~\bibnamefont{Slutsky}} \bibnamefont{et~al.},
  \bibinfo{journal}{Class. Quant. Grav.} \textbf{\bibinfo{volume}{27}},
  \bibinfo{pages}{165023} (\bibinfo{year}{2010}), \eprint{1004.0998}.

\bibitem[{\citenamefont{Abadie et~al.}(2010{\natexlab{c}})}]{S5Calibration}
\bibinfo{author}{\bibfnamefont{J.}~\bibnamefont{Abadie}} \bibnamefont{et~al.}
  (\bibinfo{collaboration}{LIGO Scientific Collaboration}),
  \bibinfo{journal}{Nuclear Instruments and Methods in Physics Research}
  \textbf{\bibinfo{volume}{A624}}, \bibinfo{pages}{223}
  (\bibinfo{year}{2010}{\natexlab{c}}), \eprint{arxiv:1007.3973}.

\bibitem[{\citenamefont{Biswas et~al.}(2009)\citenamefont{Biswas, Brady,
  Creighton, and Fairhurst}}]{Biswas:2007ni}
\bibinfo{author}{\bibfnamefont{R.}~\bibnamefont{Biswas}},
  \bibinfo{author}{\bibfnamefont{P.~R.} \bibnamefont{Brady}},
  \bibinfo{author}{\bibfnamefont{J.~D.~E.} \bibnamefont{Creighton}},
  \bibnamefont{and}
  \bibinfo{author}{\bibfnamefont{S.}~\bibnamefont{Fairhurst}},
  \bibinfo{journal}{Class. Quantum Grav.} \textbf{\bibinfo{volume}{26}},
  \bibinfo{pages}{175009} (\bibinfo{year}{2009}), \eprint{arXiv:0710.0465}.

\bibitem[{\citenamefont{Keppel}(2009)}]{Keppel:thesis}
\bibinfo{author}{\bibfnamefont{D.}~\bibnamefont{Keppel}}, Ph.D. thesis,
  \bibinfo{school}{Caltech}, \bibinfo{address}{Pasadena, CA}
  (\bibinfo{year}{2009}),
  \urlprefix\url{http://resolver.caltech.edu/CaltechETD:etd-05202009-115750}.

\bibitem[{\citenamefont{{Mandel} and
  {O'Shaughnessy}}(2010)}]{MandelOShaughnessy:2010}
\bibinfo{author}{\bibfnamefont{I.}~\bibnamefont{{Mandel}}} \bibnamefont{and}
  \bibinfo{author}{\bibfnamefont{R.}~\bibnamefont{{O'Shaughnessy}}},
  \bibinfo{journal}{Classical and Quantum Gravity}
  \textbf{\bibinfo{volume}{27}}, \bibinfo{pages}{114007}
  (\bibinfo{year}{2010}), \eprint{0912.1074}.

\bibitem[{\citenamefont{McClintock et~al.}(2006)}]{McClintock:2006xd}
\bibinfo{author}{\bibfnamefont{J.~E.} \bibnamefont{McClintock}}
  \bibnamefont{et~al.}, \bibinfo{journal}{Astrophys. J.}
  \textbf{\bibinfo{volume}{652}}, \bibinfo{pages}{518} (\bibinfo{year}{2006}),
  \eprint{astro-ph/0606076}.

\bibitem[{\citenamefont{Rodr\'iguez}(2007)}]{Rodriguez:2007}
\bibinfo{author}{\bibfnamefont{A.}~\bibnamefont{Rodr\'iguez}}, Master's thesis,
  \bibinfo{school}{Louisiana State University} (\bibinfo{year}{2007}),
  \eprint{arXiv:0802.1376}.

\end{thebibliography}

\end{document}